\documentclass[acmsmall]{acmart}
\usepackage{multicol}
\usepackage{color}
\usepackage{listings}
\usepackage{xspace}
\usepackage{multirow}
\usepackage[normalem]{ulem}
\usepackage[caption=false]{subfig}
\usepackage{array}
\usepackage{bm}
\usepackage{enumitem}
\usepackage{tabularx}

\usepackage{fancyvrb}
\newsavebox{\mylisting} % for use with lrbox

\newcommand\grumbler[3]{\textcolor{#1}{#2 says: #3}}
\newcommand\mike[1]{\grumbler{blue}{Mike}{#1}}
\newcommand\ada[1]{\grumbler{magenta}{Ada}{#1}}

\newcommand{\ifclib}{\textsc{Cocoon}\xspace}
\newcommand{\Ifclib}{\ifclib}
\newcommand{\IfcLib}{\Ifclib}
\newcommand{\nightlyversion}{1.69.0-nightly\xspace}

\definecolor{gitadd}{HTML}{00A64F}
\definecolor{gitdel}{HTML}{c94238}
\newcommand{\diffadd}[1]{\color{gitadd}{\texttt{#1}}}
\newcommand{\diffdel}[1]{\color{gitdel}{\texttt{#1}}}
\newcommand{\code}[1]{\lstinline{#1}}
\newcommand{\bench}[1]{\textsf{#1}}
\newcommand{\ignore}[1]{}

\newcommand\rewrite[1]{\rewriteparam{#1}{F}}
\newcommand\rewritenocheck[1]{\rewrite{#1}}
\newcommand\rewritesimple[1]{\rewriteparam{#1}{T}}
\newcommand\rewriteparam[2]{\ensuremath{\tau(#1, \mathit{#2})}\xspace}

\newcommand\sbscheck{\lstinline|check_ISEF|}
\newcommand\sbscheckunsafe{\lstinline|check_ISEF_unsafe|}

\definecolor{GrayCodeBlock}{RGB}{241,241,241}
\definecolor{BlackText}{RGB}{110,107,94}
\definecolor{RedTypename}{RGB}{182,86,17}
\definecolor{GreenString}{RGB}{96,172,57}
\definecolor{PurpleKeyword}{RGB}{184,84,212}
\definecolor{GrayComment}{RGB}{120,120,120}
\definecolor{GoldDocumentation}{RGB}{180,165,45}
\lstdefinelanguage{rust}
{
    columns=fullflexible,
    keepspaces=true,
    showstringspaces=false,
    frame=single,
    framesep=0pt,
    framerule=0pt,
    framexleftmargin=4pt,
    framexrightmargin=4pt,
    framextopmargin=5pt,
    framexbottommargin=3pt,
    xleftmargin=18pt,
    xrightmargin=4pt,
    backgroundcolor=\color{GrayCodeBlock},
    basicstyle=\ttfamily\color{BlackText},
    keywords={
        true,false,
        unsafe,async,await,move,
        use,pub,crate,super,self,mod,
        struct,enum,fn,const,static,let,mut,ref,type,impl,dyn,trait,where,as,
        break,continue,if,else,while,for,loop,match,return,yield,in
    },
    keywordstyle=\color{PurpleKeyword},
    ndkeywords={
        bool,u8,u16,u32,u64,u128,i8,i16,i32,i64,i128,char,str,
        Self,Option,Some,None,Result,Ok,Err,String,Box,Vec,Rc,Arc,Mutex,Cell,RefCell,HashMap,BTreeMap,
        macro_rules
    },
    ndkeywordstyle=\color{RedTypename},
    comment=[l][\color{GrayComment}\slshape]{//},
    morecomment=[s][\color{GrayComment}\slshape]{/*}{*/},
    morecomment=[l][\color{GoldDocumentation}\slshape]{///},
    morecomment=[s][\color{GoldDocumentation}\slshape]{/*!}{*/},
    morecomment=[l][\color{GoldDocumentation}\slshape]{//!},
    stringstyle=\color{GreenString},
    string=[b]",
    escapeinside={(*@}{@*)}
}

\newif\iftrim

\usepackage{pifont}

\newcommand{\tickNo}{\ding{55}}

\setcopyright{rightsretained}
\acmDOI{10.1145/3649817}
\acmYear{2024}
\copyrightyear{2024}
\acmSubmissionID{oopslaa24main-p19-p}
\acmJournal{PACMPL}
\acmVolume{8}
\acmNumber{OOPSLA1}
\acmArticle{100}
\acmMonth{4}
\received{15-OCT-2023}
\received[accepted]{2024-02-24}

\ignore{
\begin{verbatim}
gs -sDEVICE=pdfwrite -dNOPAUSE -dBATCH -dSAFER -dFirstPage=1  -dLastPage=24  -sOutputFile=submission.pdf RUST_Paper.pdf
gs -sDEVICE=pdfwrite -dNOPAUSE -dBATCH -dSAFER -dFirstPage=25 -dLastPage=999 -sOutputFile=supplement.pdf RUST_Paper.pdf
\end{verbatim}
}

\ccsdesc[500]{Security and privacy~Information flow control}
\ccsdesc[500]{Software and its engineering~Access protection}

\begin{document}
	
	\title{Cocoon: Static Information Flow Control in Rust}
	
	\author{Ada Lamba}
	\orcid{0009-0000-9605-3999}
	\affiliation{%
		\institution{Ohio State University}
		\city{Columbus}
		\country{USA}
	}
	\email{lamba.39@osu.edu}
	
	\author{Max Taylor}
	\orcid{0009-0005-7873-9694}
	\affiliation{%
		\institution{Ohio State University}
		\city{Columbus}
		\country{USA}
	}
	\email{taylor.2751@osu.edu}
	
	\author{Vincent Beardsley}
	\orcid{0000-0003-2373-7171}
	\affiliation{%
		\institution{Ohio State University}
		\city{Columbus}
		\country{USA}
	}
	\email{beardsley.49@osu.edu}
	
	\author{Jacob Bambeck}
	\orcid{0009-0007-3136-0753}
	\affiliation{%
		\institution{Ohio State University}
		\city{Columbus}
		\country{USA}
	}
	\email{bambeck.14@osu.edu}
	
	\author{Michael D. Bond}
	\orcid{0000-0002-8971-4944}
	\affiliation{%
		\institution{Ohio State University}
		\city{Columbus}
		\country{USA}
	}
	\email{mikebond@cse.ohio-state.edu}
	
	\author{Zhiqiang Lin}
	\orcid{0000-0001-6527-5994}
	\affiliation{%
		\institution{Ohio State University}
		\city{Columbus}
		\country{USA}
	}
	\email{zlin@cse.ohio-state.edu}
 
\iffalse
%\onecolumn
\pagestyle{empty}
\input{cover}
%\twocolumn
\clearpage
\pagestyle{standardpagestyle}
\setcounter{page}{1}
\fi

    \begin{abstract}

    % \medskip
    % \framebox[\textwidth]{\Large \bf Please do not distribute. This work is under double-blind review.}
    % \medskip

    Information flow control (IFC) provides confidentiality by enforcing noninterference, which ensures that high-secrecy values cannot affect low-secrecy values.
    Prior work introduces fine-grained IFC approaches that modify the programming language and use nonstandard compilation tools, impose run-time overhead, or report false secrecy leaks---all of which hinder adoption.

    This paper presents \ifclib, a Rust library for static type-based IFC that uses the unmodified Rust language and compiler.
    The key insight of \ifclib lies in leveraging Rust's type system and procedural macros to establish an effect system that enforces noninterference.
    % allows applications to safely compute arbitrary functions on secret data.
    A performance evaluation shows that using \ifclib increases compile time but has no impact on application performance.
    To demonstrate \ifclib's utility, we retrofitted two popular Rust programs, the Spotify TUI client and Mozilla's Servo browser engine, to use \ifclib to enforce limited confidentiality policies.
    % The results show that applications can be retrofitted to use \ifclib with limited modifications, at least to protect a single value, with negligible impact on run-time and compile-time performance.
    \end{abstract}

    \keywords{information flow control, type and effect systems, Rust}
    \maketitle

    %\mike{General comment: Let's say ``application'' (or maybe ``application developer/programmer'' in some cases) instead of ``user'' when talking about the entity using \ifclib.}

    \ignore{
    \mike{Writing consistency (assuming American English including American typography) and latex stuff:
    \begin{itemize}
    \item ``Adverb adjective noun'' shouldn't have a hyphen, e.g., ``widely used software.'' Only ``adjective-noun noun'' and ``adjective-adjective noun'' should have hyphens.
    \item Footnotes generally go after punctuation, not before.
    \item ``That'' vs.\ ``which'': In a nutshell, use ``that'' instead of ``which'' when either seems okay (\url{https://www.lexico.com/grammar/that-or-which}, \url{https://www.grammarly.com/blog/which-vs-that/}).
    \end{itemize}}
    }

    % what are we doing and who cares
    \section{Introduction}
\label{sec:intro}

Confidentiality is a fundamental property of secure computing systems.
Confidentiality, or secrecy, is usually achieved through \emph{access control}, which regulates \emph{which} entities may access \emph{which} data, but \emph{not} what happens to the data after it is accessed, resulting in a large trusted computing base.
As a robust alternative to access control, \emph{information flow control (IFC)} ensures that high-secrecy values cannot flow to (i.e., cannot affect) low-secrecy values~\cite{keeping_secrets,denning_lattices}.\footnote{IFC generally ensures not only confidentiality but also integrity, by ensuring that low-integrity values cannot flow to high-integrity values, but this paper handles only confidentiality.}
% JFlow,SM_03.

IFC is especially valuable as a \emph{fine-grained} mechanism that protects the flow of data values within applications, ensuring they do not misuse privileges. Consider, for example, a virus checker application with privileges to both read data from the user's disk and send data to a remote server (to request software updates). However, the virus checker should \emph{not} be allowed to send the contents of the user's disk to the remote server---a confidentiality policy that fine-grained IFC can enforce.
% \maxt{Can we use a more relevant example from Spotify TUI or Servo?}
% \mike{Maybe we could, but the security policies this paper actually evaluates for Spotify TUI and Servo don't seem particularly elucidating or compelling.}
As a result, the virus checker does not need to be trusted by the user.
In general, fine-grained IFC reduces the trusted computing base from the entire application to the relatively few operations that explicitly violate IFC by \emph{declassifying} secret values.

Fine-grained IFC can be supported either dynamically at run time, which incurs run-time costs~\cite{AF_2009, laminar, coInflow},
% (\autoref{subsec:related}),
or statically at compile time.
%\emph{Dynamic IFC} tracks and checks flows at run time~\cite{AF_2009, laminar, coInflow}, incurring run-time overhead.
%\mike{Hmm, dynamic IFC has the same ``doesn't work for unmodified mainstream imperative languages'' problem as static type-based analysis, right?} \ada{The sources we cite extend the language, so yes, but I don't know that that is universal.}
% Furthermore, dynamic analysis has inherent difficulty handling \emph{implicit} information flows (e.g., \lstinline$if (secret_var) { non_secret_var = 1; }$), and it can introduce unexpected failures and termination channels.
% \mike{We agree with reviewer that this isn't really true.}
%In contrast to dynamic IFC, static IFC incurs no run-time overhead, by performing IFC checks at compile time.
% can handle implicit flows, and can avoid termination channels.
Static IFC can use whole-program analysis or modular type-based analysis.
% \maxt{Do we want to reword this to distinguish between external static analysis tools and static typing? This was mentioned in our reviews.}
% \mike{Reworded}
\label{subsec:intro-static-analysis}%
\emph{Whole-program static analysis} conservatively models the flow of values through expressions and variables and ensures that high-secrecy values do not flow to low-secrecy sinks~\cite{BBBPRR_17, SAILS, HS_12, JavaPDGs}.
However, whole-program analysis is non-compositional, so it scales poorly and supports incremental deployment poorly.
% Static analysis approaches require the programmer to annotate the secrecy label for each new secret variable, but not the results of operating on these variables. Then, at compile time, the analyzer determines the required secrecy level for all effects of the secret variables and performs IFC checks.
%(though this, and many other works, do not).
% Furthermore, static analysis generally employs whole-program analysis and is thus implemented separately from the compiler, adding to the tools that must be maintained and trusted.
A recent analysis called \emph{Flowistry} leverages Rust ownership constraints to compute relatively precise function summaries without analyzing the functions' bodies~\cite{flowistry}.
% However, Flowistry is an intraprocedural analysis and reports false flows (\autoref{subsec:related}).
In contrast, \emph{static type-based analysis} uses types to represent the security labels of variables and expressions, and a correctly typed program guarantees IFC~\cite{MAC, DepSec, SM_03, JFlow, Jif, ML_97, JRIF, chong_dissertation, FlowCAML, spark, IFC_multithreaded, SV_type_model}.
While type-based analysis allows for modular checking and incremental deployment, existing solutions rely
on a \emph{modified} language and compiler,\footnote{An exception is type-based IFC analysis in Haskell (\autoref{subsec:related}).}
impeding adoption.
% ~\cite{SM_03, JFlow, Jif, ML_97, JRIF, chong_dissertation, FlowCAML, spark, IFC_multithreaded, SV_type_model}.
% \sout{Keeping the modifications up to date, and programmers cannot use standard development tools.} \mike{As long as we say this elsewhere let's remove it here.}
% has several drawbacks: Keeping the modified language up-to-date requires significant effort; programmers must trust the modifications; and programmers cannot use standard development tools such as IDE features and debuggers.
% \autoref{subsec:related} covers prior work in more detail.

Why must prior work on type-based analysis \emph{modify} the language and compiler?
Because the language's type system is not expressive enough to ensure noninterference statically.
% \mike{I think the rest of this paragraph may be too much detail here, and should be moved to a later section.}
% \ada{I think this needs to be introduced here (or *maybe* at the beginning of contributions). We discuss an ``effect system'' throughout the paper so we need to define it somewhere and the point of this paragraph is to point out that prior work hasn't really addressed effect systems for languages like Rust; that fits in here. So, I'm not sure where else it would make sense to put it.}
% \mike{Makes sense. Revised some.}
The key technical challenge is how to allow programs to define and use functions on secret data without permitting unchecked side effects that might violate noninterference. In theory, an \emph{effect system} for encoding the side effects of a computation would constrain the behaviors of functions on secret data~\cite{effect-systems}. While effect systems have been implemented in \emph{Haskell} for type-based IFC~\cite{MAC, DepSec, dependency-core-calculus},
% \maxt{We should probably also cite the calculus of dependency paper here.}
% \mike{Addressed.}
effect systems are not supported by mainstream imperative languages including Java, C, C++, and Rust.
% \mike{Do we also cover this later? We at least don't cite \cite{effect-systems} in any other section. Move/incorporate this paragraph's content to a later section?}

\subsection*{Contributions}

This paper introduces \emph{\ifclib}, the first static type-based IFC for a mainstream imperative language. \Ifclib is a Rust library for the unmodified Rust language that works with the standard Rust compiler (although \ifclib relies on some Rust features that are not yet part of stable Rust). \Ifclib ensures that a program compiles only if
% that identifies secret data correctly is able to compile,
it does not leak secrets, except by explicitly \emph{declassifying} secrets or explicitly using unsafe Rust code.
\Ifclib provides a rudimentary effect system for noninterference that limits computations' side effects using a novel approach leveraging Rust
features including ownership, traits,
% \maxt{``expressive typing'' is vague -- perhaps we mean auto traits and the like? Every other phrase describes a specific technology. I think we should match.}
% \mike{Revised}
and procedural macros.
% While Rust provides features unique to mainstream imperative languages---such as ownership and lifetime features, an expressive type system, and procedural macros---it is hardly obvious how to use them to create an effect system. We leverage Rust's type system in a novel way to create an effect system that limits the side effects a computation can have.

% This work focuses on supporting secrecy labels for language-level static IFC.
% This paper's key insight lies in how \ifclib allows applications to express arbitrary operations on secrets by providing an effect system that limits the side effects that functions have on secret data.
% While Rust provides features unique to mainstream imperative languages---such as ownership and lifetime features, an expressive type system, and procedural macros---it is hardly obvious how to use them to create an effect system. We leverage Rust's type system in a novel way to create an effect system that limits the side effects a computation can have.

% Our prototype implementation of \ifclib is a Rust library that provides secret types and facilities for performing arbitrary secure operations on secret values. Importantly, \ifclib differs from prior type-based approaches in that it works with the unmodified Rust language and compiler, i.e., the modified applications are pure Rust programs.
We evaluated \ifclib's effectiveness and efficiency by retrofitting two popular Rust applications, a Spotify client and the Servo web browser engine, to enforce a confidentiality policy for a single secret value in each program. These case studies demonstrate that \ifclib's approach works in real applications and can be incrementally deployed.
% While the program modifications required can be significant, they are generally limited to parts of the program that deal with secret values.
% %
The evaluation also shows \ifclib does not affect application run time or memory usage, although it can affect compile time if a substantial fraction of the program code handles secret values.

\Ifclib thus advances the state of the art in fine-grained IFC support by demonstrating a novel approach that requires no language or compiler modifications.
% \Ifclib can be incrementally deployed on existing systems, resulting in applications with a small trusted codebase.
% The key enabling insight is providing an effect system that captures whether Rust functions have side effects. As a result, programs written in a mainstream, high-performance language can be trusted (after being audited for a few special operations) not to leak secret data.
The current design has several limitations, including supporting only static secrecy labels and placing restrictions on the programming model
% Enhancements such as eliminating programming model limitations, reducing the type annotation burden, and supporting integrity labels and dynamic labels
% and integrating with OS-level IFC
% are outside this paper's scope and relegated to future work
(\autoref{subsec:limitations}).

This paper makes the following contributions:
\begin{itemize}[leftmargin=*]
    \item a fine-grained IFC approach that is the first to provide static type-based IFC for a mainstream imperative language without language or compiler modifications;
    \item novel support for a rudimentary effect system that captures side effects that could leak secrets;
    \item two case studies on real Rust programs that demonstrate that \ifclib works for real programs and adds negligible or nonexistent run-time and compile-time overheads; and
    \item a performance evaluation showing that \ifclib has no detectable run-time impact, and its impact on compile time is commensurate with how much code that deals with secret values.
    % when \ifclib is applied to entire programs, it impacts compile time significantly,

\end{itemize}

    \section{Motivating Example}
\label{sec:motivating-example}

%\mike{This section has turned out to be entirely example driven---it's not really an ``overview''---so I've titled it ``Motivating Example.''}

\autoref{fig:insecure-cal} shows example Rust code that computes the number of overlapping days of availability on two calendars, belonging to Alice and Bob. Each calendar is a map from a day of the week, represented as a string, to a boolean indicating whether Alice or Bob is available on that day. Suppose that Alice and Bob's calendars are secret. Specifically, the values (booleans) in their calendars are secrets, although the keys (strings) are not. For simplicity we assume that the calendars have identical key sets.
The example computes and reveals only the number of overlapping days in Alice and Bob's availability.
% without otherwise revealing their schedules externally.

The syntax \lstinline{for (day, available) in alice_cal} iterates over all key--value pairs in Alice's calendar.
% which is just a map from string names (like days of the week) to boolean values indicating availability.
Note that the types of \lstinline{day} and \lstinline{available} are \lstinline{&String} and \lstinline{&bool}, meaning that they are actually (immutable) \emph{references} to the values. The expression \lstinline{bob_cal.get(&day)} calls \lstinline{HashMap::get(&self, }\allowbreak\lstinline{&String) -> Option<&bool>}, which is a method that takes a reference to a \lstinline{String} as a parameter. Rust allows an unlimited number of immutable references, \emph{or} a single mutable reference, to a value at a time.
% \maxt{It should be a reference to an object, right? Values can't be referenced?}
% \ada{Agreed}
% \mike{Disagree. Everything in Rust is a value, not an object.}
The \lstinline{get} method returns a value of type \lstinline{Option<&bool>}, which represents a reference to a boolean, or an error if the key was not found in \lstinline{bob_cal}. The call to \lstinline{Option::unwrap(self) -> &bool} returns the reference to the inner boolean, or fails otherwise.
% In this example, we assume that the \emph{keys} in Alice and Bob's calendars are the same and are \emph{not} supposed to be secret.
Since \lstinline{unwrap()} returns type \lstinline{&bool} (i.e., reference to a boolean), the code uses the \lstinline{*} operator to dereference (and implicitly copy) \lstinline{unwrap()}'s return value.

To ensure that the code in \autoref{fig:insecure-cal} does not leak more information about Alice or Bob's schedule than the number of overlapping days, one must trust the code or audit it carefully.
% %
Alternatively, fine-grained IFC
% \emph{information flow control (IFC)}
provides a way to \emph{ensure} that code cannot leak secrets to untrusted entities, by disallowing high-secrecy values from flowing to low-secrecy values.
% ~\cite{keeping_secrets, denning_lattices}.
% IFC can ensure both \textit{confidentiality} (secrecy) and \textit{integrity};
%\mike{FYI I removed mention of the the duality and removed two citations, \cite{B_73} and \cite{BN_04}, which were the only citations to those articles. Do we want to cite them somewhere?} \ada{No, we can remove them.}
% A duality exists between these two, as confidentiality requires that data does not flow into an unintended sink, while integrity requires that data does not flow out of an unintended source~\cite{B_73, BN_04}.
% \ignore{Said another way, confidentiality is concerned with what information is leaked from a program while integrity is concerned with the corruption of data.}%
% for simplicity, this paper supports confidentiality but not integrity.

Information flows can be explicit or implicit. In an \emph{explicit flow},
% a value flows directly to another value, i.e.,
a data dependency exists between the two values.
In \autoref{fig:insecure-cal} the return value of \lstinline{bob_cal.get(&day)} flows to \lstinline{Option::unwrap()}'s parameter via an explicit flow.
% The left side of \autoref{fig:flows} shows an explicit flow: \lstinline{sec} is written directly to another variable, which is then sent to an untrusted socket. This example leaks the value of \lstinline{sec}, which the socket receiver can determine by subtracting 2 from the value.
An \textit{implicit flow} is a flow that involves control dependence.
% one in which the the control flow of the program can be used to infer information about secret data~\cite{KHHJ_2008}, i.e., the data dependence involves a control dependency.
In \autoref{fig:insecure-cal}, an implicit flow exists from \lstinline{available}'s value (and from the return value of \lstinline{bob_cal.get(&day)}) to \lstinline{count}'s value.
% The right side of \autoref{fig:flows} shows an implicit flow. This example constitutes a leak because information about \lstinline{sec}'s value (whether it is 42) can be inferred from the value received by the socket.
% \maxt{Leading with an example is good. Is there a concrete definition of an implicit flow we can now provide here?}
% \mike{Addressed with control/data dependence}

\iffalse
\begin{figure}[t]
    \begin{minipage}[t]{.47\textwidth}
\begin{lstlisting}
(*@\label{lst:explicit flow}@*)let sec: i32 = stdin().read().parse();
let another_value = sec + 2;
// ...
tcp_socket.send(another_value);
\end{lstlisting}
    \centering \small \textsf{Explicit flow}
    \end{minipage}
    \hfill
    \begin{minipage}[t]{.47\textwidth}
\begin{lstlisting}
(*@\label{lst:implicit flow}@*)let sec: i32 = stdin().read().parse();
// ...
if sec == 42 { tcp_socket.send(1); }
else         { tcp_socket.send(0); }
\end{lstlisting}
    \centering \small \textsf{Implicit flow}
    \end{minipage}
    \caption{Examples of an explicit flow (left) and an implicit flow (right) of a secret value to an untrusted entity.
    \mike{Let's talk about illegal flows rather than untrusted entities.}
    \ignore{The syntax for reading an integer from the terminal is simplified for illustrative purposes.}}
    \label{fig:flows}
\end{figure}
\fi

\begin{figure}
\begin{lstlisting}
let alice_cal: HashMap<String, bool> = /* { "Monday" -> true, ... } */
let bob_cal:   HashMap<String, bool> = /* { "Monday" -> false, ...} */
let mut count = 0;
for (day, available) in alice_cal {
    if available && *bob_cal.get(&day).unwrap() {
        count += 1;
    }
}
println!("Overlapping days: {}", count);
\end{lstlisting}
\caption{Rust code that computes the overlapping number of days in two calendars.}
\label{fig:insecure-cal}
\bigskip\bigskip
% \end{figure}
% \begin{figure}
\begin{lstlisting}[escapechar=@,numbers=left]
let alice_cal: HashMap<String, Secret<lat::A,bool>> = /* { "Monday" -> Secret(true), ... } */
let bob_cal:   HashMap<String, Secret<lat::B,bool>> = /* { "Monday" -> Secret(false), ...} */
let mut count = secret_block!(lat::AB { wrap_secret(0) }); @\label{line:secure-cal:first-secret-block}@
for (day, available) in alice_cal {
    secret_block!(lat::AB { @\label{line:secure-cal:begin-second-secret-block}@
        if unwrap_secret(available) &&
           *unwrap_secret_ref(::std::option::Option::unwrap(
               ::std::collections::HashMap::get(&bob_cal, &day))) {
            *unwrap_secret_mut_ref(&mut count) += 1;
        }
    }); @\label{line:secure-cal:end-second-secret-block}@
}
println!("Overlapping days: {}", count.declassify()); @\label{line:secure-cal:declassify}@
\end{lstlisting}
\caption{Rust code that has the same functionality as \autoref{fig:insecure-cal} but uses \ifclib to ensure IFC.}
\label{fig:secure-cal}
\end{figure}

Consider
\autoref{fig:secure-cal}, which has the same functionality as \autoref{fig:insecure-cal}, but it ensures IFC using this paper's approach, \ifclib. The booleans in Alice and Bob's calendars are wrapped in a \lstinline{Secret} type and have labels $\{ a \}$ and $\{ b \}$, where $a$ and $b$ are \emph{policies} (also called \emph{tags}) corresponding to secrecy privileges of Alice and Bob, respectively. A label $L_1$ is \emph{at least as secret as} label $L_2$ if and only if $L_1 \supseteq L_2$.

Applications can read and write secret values only inside of lexically scoped regions called \emph{secret blocks}, denoted by the \lstinline{secret_block!}\footnote{In Rust, all macro names end with the \lstinline{!} character.} macro call. Before the loop, the code uses a secret block to create a new secret value \lstinline{0} (zero) with label $\{ a, b \}$, which is assigned to \lstinline{count}. The secret block inside the loop\footnote{One could instead wrap the whole loop in a single secret block, but in the spirit of minimizing code in secret blocks---and to showcase more of \ifclib's programming model---the code wraps only the loop's body in a secret block.} has label $\{ a, b \}$; \ifclib ensures that the block's inputs must be no more secret than $\{ a, b \}$, and its output(s) must be no less secret than $\{ a, b \}$, thus preventing illegal flows.

The block uses calls to \lstinline{unwrap_secret}, \lstinline{unwrap_secret_ref}, and \lstinline{unwrap_secret_mut_ref} to access secret values. \Ifclib ensures that these calls are only allowed if the unwrapped value's label is no more secret than the secret block's label. \lstinline{unwrap_secret_mut_ref()} gets \emph{mutable} access to a secret value, allowing both read and write access, requiring the value to have the \emph{same} label as the secret block.

The secret block uses fully qualified calls to Rust Standard Library functions
% \lstinline{HashMap::get()} and \lstinline{Option::wrap()}
(e.g., \lstinline{::std::option::}\allowbreak\lstinline{Option::unwrap(...)}), which \ifclib requires
% for calls to Rust Standard Library functions,
in order to check that the callee cannot leak secrets.

After the loop completes, the program explicitly declassifies the secret value \lstinline|count| in order to print its value.
% A call to \lstinline|declassify()| returns the value stored in the given \lstinline|Secret|.
Declassifications are part of the trusted computing base and must be audited.

    \section{Requirements and Assumptions}
\label{sec:requirements-and-assumptions}

Fine-grained information flow control (IFC) is not used widely in practice because it is not practical.
% A key reason is that existing language-level IFC approaches are impractical.
We argue that a practical solution must provide IFC for programs written in a mainstream imperative language without modifying the language or its compiler.
% \sout{ or incurring run-time overhead}.
% \mike{Incurring run-time overhead is unavoidable in general---i.e., for dynamic IFC.}
% \mike{This section should motivate the goals as deriving from the problems and described in the previous sections---we didn't come up with them ourselves.}
% Our solution fundamentally has two goals: to provide static IFC and to do so without requiring additional programming tools. The purpose of this section is to formalize these requirements and provide intuition behind how \ifclib will meet these requirements.
% The rest of this section describes these requirements in more detail.

\subsection{Assumptions and Nonrequirements}
\label{subsec:assumptions}

This paper focuses on achieving fine-grained IFC for static secrecy labels using an off-the-shelf language and compiler. This work does \emph{not} aim to support integrity labels, dynamic labels,
% IFC checking within an unsafe code block, \ada{why is this commented out? We DON'T check unsafe code.}
or OS integration.\footnote{Since OS integration is outside our scope, the programming model assumes that every OS-level entity is an untrusted entity. It is the programmer's responsibility to identify secrets coming from OS-level entities such as files and sockets, and the program must declassify a secret value before sending it to any OS-level entity.}
% Our work aims to provide \emph{termination-insensitive noninterference} (explained below);
% IFC violations due to programmer error or malice,
It does not protect side channels such as termination and timing channels. Finally, although the programming model is pure Rust, it has several restrictions, a challenge we leave for future work.
% \maxt{We should state ``termination-insensitive IFC'' here. This terminology likely needs discussed in background.}
% \mike{Mentioned here and added brief explanation below.}
% %
% since all OS-level entities are considered to be untrusted sinks.
% Thus the program must identify secret values coming from OS-level entities. And the program cannot write secret values to OS-level entities---it must declassify the values first.
\autoref{subsec:security-guarantees-threats} and \autoref{subsec:limitations} describe \ifclib's guarantees and limitations in more detail.

Our work aims to provide \emph{termination-insensitive noninterference} (explained below),
but we do not present a proof of correctness, which would be a notable research contribution by itself. Other work has presented correctness proofs, but only for small, toy languages~\cite{confidence_integrity_untrustedhosts, robust_declassficiation, SIF, curricle}. As Zheng and Myers point out, ``[p]roofs for noninterference exist for numerous security-typed languages, but not for any language as expressive as Jif''~\cite{SIF}.
% IFC solutions for larger-scale languages do not include correctness proofs~\cite{JFlow, SafeIFC_DecentralizedLabels, SIF}, with
The authors of Viaduct (a compiler for secure distributed programs) note that ``a full correctness proof for the Viaduct compiler would be a significant research achievement \ldots''~\cite{viaduct}.
%\ada{Mention Curricle~\cite{curricle} here? They give a proof for their formal model but not the Rust implementation of said model.}
%\mike{Added citation to Curricle for ``small, toy languages'' above.}

%Untrusted entities (e.g., untrusted sockets or processes) are likewise identified.
%\ada{This sentence implies to me that either we are pointing out unsafe entities or the user is. Is this in the paper or done by \ifclib? It also seems to me that this a little vague: does this mean we point out any possible unsafe entity or only entities that are unsafe in the context of their use? Is this identification through our error messages?}
%\mike{They're identified by application programmers or as part of integration with an OS-level IFC approach, e.g., the OS-level IFC approach tracks policies (i.e., permissions) for sockets, files, processes, users, etc.}

\subsection{Requirements}
\label{subsec:requirements}

% This paper aims to provide IFC support through a Rust library, without modifying the Rust language or compiler.
Here we motivate and describe the requirements that our IFC approach needs to meet.

\subsubsection*{Widely Used, High-Performance Language}

A drawback of many prior static IFC solutions is that they require the use of an academic language that inhibits their adoption (\autoref{subsec:related}). Thus, we aim to develop an IFC solution that uses a popular, high-performance language: \emph{Rust}. Rust was designed by Graydon Hoare at Mozilla Research in 2010 and had its first stable release in 2015~\cite{Hoare_Rust_Blog, Mozilla_Rust}.
% ~\cite{Rust_GitHub}.
Since then it has had 95 releases, over 3,500 contributors, and has held the title of Stack Overflow's ``most loved language'' by developers every year since 2016~\cite{stackOverflow_survey}.
Rust is commonly used in large-scale applications; for example, many core features of Mozilla Firefox are primarily written in Rust~\cite{rust_production_users}.

\subsubsection*{Off-the-Shelf Language and Compiler Support}

In contrast to prior solutions that extend a language and modify the compiler or rely on a custom compiler (\autoref{subsec:related}),
% limiting practical applicability,
we aim to enforce IFC without requiring a customized language or compilation tools.
%The motivation for this goal is to reduce programmer burden.
%\mike{This problem doesn't seem like ``programmer burden'' exactly. I suggest just removing the sentence, since we already mentioned ``practical applicability'' above.}
% As discussed above, many IFC solutions require using a limited, academic language, whole-program analysis, or using a separate compilation tool. This can make integration into existing systems difficult and require additional tools or skills from the programmer. Satisfying our second goal then allows for easier utilization of the library and less overhead during compile-time.
In the context of the Rust language, we aim to provide IFC entirely through a library. Programs are written in unmodified Rust using the IFC library and compiled with the unmodified Rust compiler, although our design relies on some language features that are not yet part of stable Rust.
% Thus we aim to provide IFC entirely in the unmodified ---specifically In summary, the second goal requires that we leverage pure Rust syntax (e.g., macros, closures) as well as the the guarantees that the RustC compiler provides to complete our checks.
The library is trusted (i.e., is part of the trusted codebase) along with the Rust compiler and standard library. The application is untrusted, except for \lstinline{unsafe} and declassify operations, which must be trusted or audited.
In Rust, \lstinline{unsafe} operations already forgo guarantees of memory and type safety.

\subsubsection*{Soundness}
\label{subsec:goals-soundness}

We aim to
provide \emph{termination-insensitive noninterference}: In the absence of explicit declassification or explicitly \lstinline{unsafe} code, high-secrecy values cannot affect low-secrecy values but may affect whether the program terminates~\cite{dependency-core-calculus,termination-insensitive-noninterference-leaks}.
% Otherwise the program will not compile.
% \maxt{We don't have a notion of secrecy applied to output channels. To be precise, perhaps we should say that ``high-secrecy values cannot affect low-secrecy values.'' I also suggest qualifying this statement: \emph{in the absence of an explicit declassify or unsafe block.}}
% \mike{Applied.}
This model handles explicit and implicit flows, but it does not address side channels created by timing or microarchitectural states.
% nor by concurrency interleavings (cf. Figure~10 in \cite{laminar-toplas-2014}).
% \mike{I guess that's a timing side channel.}

% A program in which a secret value can flow to a less-secret value (via explicit or implicit flows) should not compile successfully unless the flow is accomplished by explicitly \emph{declassifying} secret values or by using unsafe Rust (by using the \lstinline{unsafe} keyword). Declassification and unsafe Rust are rare by design~\cite{unsafe-rust}, minimizing the parts of the application that must be trusted or audited to ensure correctness.

\subsubsection*{Incremental Deployment}

A goal is for the library to be incrementally deployable: Developers only need to modify parts of their code that deal with secret values.
% without modifying or affecting the functionality of the rest of the code.
These modifications may still be time consuming and burdensome, a challenge we leave for future work.

\subsubsection*{Run-Time Performance}

% Especially considering that Rust is a systems language that developers often choose for performance reasons,
Programs using the IFC library should perform indistinguishably from otherwise equivalent programs that do not. The Rust compiler should be able to ``optimize away'' the typing and scaffolding introduced by the application's use of the IFC library, reducing the compiled code so it performs the same as an otherwise equivalent program.
% not using the IFC library.

    % what guarantees do we provide and how does the core functionality of our solution work; how can we leverage Rust to provide these guarantees in an IFC library
    \section{\IfcLib: Static Information Flow Control in Rust}
\label{sec:design-impl}

% \ada{TODO: Stopped here converting ref to autoref - choice was arbitrary as it seemed like there were fewer refs to change than autorefs.}
% \mike{I finished the conversion.}

\Ifclib is a static IFC approach that meets \autoref{subsec:requirements}'s requirements.
\autoref{subsec:programming-model} explains how application developers use \ifclib to ensure noninterference.
\autoref{subsec:ensure-noninterference} describes how \ifclib ensures that the program will not compile unless it enforces noninterference, and \autoref{subsec:impl-details} gives implementation details.
\autoref{subsec:security-guarantees-threats} and \autoref{subsec:limitations} discuss guarantees, threats, and limitations.

% outlined in the previous section.
% It allows explicit and implicit flows while disallowing programs with illegal flows.
% It is implemented as a Rust library, and Rust applications can use it without needing a new compiler or other custom tools.
% Our evaluation shows that using \ifclib impacts compile time but does not detectably impact application run time. \maxt{I think it's too early to mention our evaluation results here -- let's just discuss the design of Cocoon.} \mike{Okay.}

% \Ifclib introduces a generic datatype for secret values,
% and it employs Rust's type system to ensure that secret values cannot flow to less-secret expressions or variables. \Ifclib permits legal application-defined flows, including implicit flows, by leveraging Rust's features
% to provide a rudimentary effect system that prohibits side effects that can leak secret values.

\subsection{\IfcLib's Programming Model}
\label{subsec:programming-model}

An application uses \ifclib's programming model to represent secret values and secrecy labels, to access secret values, and to declassify secret values.
\autoref{fig:programming-model} overviews the programming model, and
\autoref{fig:secure-cal} (introduced in \autoref{sec:motivating-example})
% (page~\pageref{fig:secure-cal})
shows example application code that uses most elements of the programming model.

% To work with secret values, an application must use \ifclib's programming model.
% \autoref{fig:programming-model} overviews the programming model: how secret values and secrecy labels are represented (\ref{subfig:model:secret-type}--\ref{subfig:model:labels}), how to access secret values (\ref{subfig:model:secret-blocks}--\ref{subfig:model:derive-isef}), and how to declassify secret values (\ref{subfig:model:declassify}).
%\mike{Also refer to \autoref{fig:secure-cal} (page~\pageref{fig:secure-cal}) in this subsection.}
%
% \paragraph{Example}
%
% \ada{What are we discussing here beyond what is given at the end of the motivating example section?}

\begin{figure}
\small
\lstset{basicstyle=\footnotesize\ttfamily}

\begin{lrbox}{\mylisting}
\begin{minipage}{\linewidth}
\centering
\begin{lstlisting}[linewidth=\textwidth]
struct Secret<Type: SecretValueSafe, SecrecyLabel: Label>
\end{lstlisting}
\end{minipage}
\end{lrbox}

\subfloat[Secret expressions and variables with non-empty secrecy have type \lstinline{Secret<T,L>}. The trait \lstinline{SecretValueSafe}, defined in \autoref{sec:application-defined-functions-visible}, restricts what values can be wrapped in \lstinline{Secret}.]{
    \usebox{\mylisting}
    \label{subfig:model:secret-type}
}\\[2.25em]

\subfloat[All possible labels for two policies, $a$ and $b$, and traits defining a partial order of labels]{
\begin{tabular}{l|p{4in}}
\textbf{Struct} & \textbf{Traits implementing struct} \\\hline
\lstinline|Label_Empty| & \lstinline|Label|, \lstinline|MoreSecretThan<Label_Empty>| \\\hline
\lstinline|Label_A| & \lstinline|Label|, \lstinline|MoreSecretThan<Label_Empty>|, \lstinline|MoreSecretThan<Label_A>| \\\hline
\lstinline|Label_B| & \lstinline|Label|, \lstinline|MoreSecretThan<Label_Empty>|, \lstinline|MoreSecretThan<Label_B>| \\\hline
\lstinline|Label_AB| & \raggedright \lstinline|Label|, \lstinline|MoreSecretThan<Label_Empty>|, \lstinline|MoreSecretThan<Label_A>|, \lstinline|MoreSecretThan<Label_B>|, \lstinline|MoreSecretThan<Label_AB>|
\end{tabular}
\label{subfig:model:labels}
}\\[2.25em]

\begin{lrbox}{\mylisting}
\begin{minipage}{\linewidth}
\centering
\begin{lstlisting}
secret_block!($\mathit{label}$ { $\mathit{body}$ })
\end{lstlisting}
\end{minipage}
\end{lrbox}

\subfloat[Secret block, which executes $\mathit{body}$ with secrecy label $\mathit{label}$]{
% \ada{Mike: Can you fix this formatting?}]{
\usebox{\mylisting}
\label{subfig:model:secret-blocks}
}\\[2.25em]

\subfloat[Operations on \lstinline{Secret} values in a secret block with secrecy label $L_B$]{
\begin{tabular}{l|l|l}
 & & \textbf{Type} \\
\textbf{Operation} & \textbf{Evaluates to} & \textbf{constraint} \\\hline

\lstinline|unwrap_secret($v$: Secret<T,$L$>) -> T| &
Inner value &
$L \subseteq L_B$ \\\hline

\lstinline|unwrap_secret_ref($v$: \&Secret<T,$L$>) -> \&T| &
Immutable ref.\ to inner value &
$L \subseteq L_B$ \\\hline

\lstinline|unwrap_secret_mut_ref($v$: \&mut Secret<T,$L$>) -> \&mut T| &
Mutable ref.\ to inner value &
$L = L_B$ \\\hline

\lstinline|wrap_secret(|$v$\lstinline|)| &
\lstinline|Secret<_,$L_B$>::new($v$)| &
N/A \\

\end{tabular}

\label{subfig:model:unwrap}
}\\[2.25em]

% \maxt{Would it be clearer if we wrote the ``L = block's label'' as a type constraint for \lstinline{wrap_secret?}}
% \ada{Agreed}
% \mike{But it's not really a type constraint, right?}
% \mike{Addressed by adding $L_B$}

\begin{lrbox}{\mylisting}
\begin{minipage}{\linewidth}
\begin{lstlisting}
#[side_effect_free_attr]
fn $\mathit{func}$($\mathit{params}$) -> $\mathit{retType}$ { $\mathit{body}$ }
\end{lstlisting}
\end{minipage}
\end{lrbox}

\subfloat[Annotation for side-effect-free functions. Secret blocks may (transitively) call only side-effect-free functions.]{
    \usebox{\mylisting}
    \label{subfig:model:side-effect-free-attr}
}\\[2.25em]

\begin{lrbox}{\mylisting}
\begin{minipage}{\linewidth}
\begin{lstlisting}
#[derive(InvisibleSideEffectFree)]
struct $s$ { $\dots$ }
\end{lstlisting}
\end{minipage}
\end{lrbox}

\subfloat[Annotation for application types that are used in secret blocks]{
    \usebox{\mylisting}
    \label{subfig:model:derive-isef}
}\\[2.25em]

\subfloat[Declassify operations on \lstinline{Secret} values]{
\begin{tabular}{l|l}
\bf Operation & \bf Evaluates to \\\hline

\lstinline|Secret<T,$L$>::declassify(self) -> T| &
Inner value \\\hline

\lstinline|Secret<T,$L$>::declassify_ref(\&self) -> \&T| &
Immutable reference to inner value \\\hline

\lstinline|Secret<T,$L$>::declassify_ref_mut(\&mut self) -> \&mut T| &
Mutable reference to inner value

\end{tabular}
\label{subfig:model:declassify}
}\\[1.5em]

\caption{\Ifclib's programming model.}
\label{fig:programming-model}
\end{figure}

\subsubsection*{Secret Values and Secrecy Labels}

Our work follows the decentralized IFC model introduced by Myers and Liskov~\citeyearpar{ML_97}.
% and extended in the literature (e.g.,~\cite{JFlow,Jif,flume,laminar}).
A secrecy \emph{label}, which describes the permissions for data, is a set of secrecy \emph{policies} (also called \emph{tags}). For example, a label $L = \{a, b\}$ contains the policies $a$ and $b$. These policies correspond to entities with different secrecy privileges, such as users of a system. Labels form a partial order on these policies, in which
% $\bot = \emptyset$ (no secrecy), $\top$ is the set of all policies, and
label $L_1$ is \emph{at least as secret as} label $L_2$ if and only if $L_1 \supseteq L_2$.
% Under IFC, values are only permitted to flow to values and entities that are at least as secret.

Every expression and variable has a static secrecy label determined by its static type. By default, expressions and variables have the empty secrecy label (lowest secrecy). An expression or variable with higher secrecy is represented by wrapping it in a \ifclib-provided type \lstinline{Secret<Type,Label>}, presented in \autoref{subfig:model:secret-type}.
\lstinline{Secret} values are opaque and must be accessed through an interface provided by \ifclib, discussed shortly.
Note that wrapping values in \lstinline{Secret} incurs no run-time cost; a \lstinline{Secret<Type,Label>} instance has the same run-time representation as a \lstinline{Type} instance.

% In \ifclib, secrecy labels are represented as static types.
% \maxt{It might be a good idea to remind the reader what a policy is -- it is defined in section 1. We should introduce a precise definition here.}
% \mike{Moved to previous paragraph.}
\autoref{subfig:model:labels} shows that labels are implemented as structs, and trait implementations\footnote{Rust's traits are analogous to Java's interfaces and C++'s abstract methods, but traits allow more expressive type constraints.} help enforce legal flows.
% \maxt{The phrase ``trait implementations'' is used but is not defined -- see comment about line \textcolor{red}{62}. Maybe mention trait earlier?}
% \mike{This is the first time we talk about traits---except to mention them in a list of Rust features we levereage. Added a footnote here.}
Specifically, a trait \lstinline{MoreSecretThan<$M$>} is implemented for label $L$ if and only if $M \subseteq L$.

In \autoref{fig:secure-cal}, \lstinline{alice_cal} and \lstinline{bob_cal} are hash maps whose values are secret boolean values with labels $\{ a \}$ and $\{ b \}$, respectively.

%A lattice structure, as introduced by Dennings~\cite{D_76}, is comprised of a partially ordered set with (at a minimum) operators for determining the least upper bound and greatest lower bound of elements in this set. Thus, this lattice defines a type of (not necessarily linear) hierarchical relationship on the set elements. Examples of a lattice are shown in Figure \ref{fig:lattices}. Visually, we want to restrict information flow such that secret information can only flow up the lattice.

%In essence, this trait really encodes the concept ``at least as secret as''. Thus, each secrecy level is more secret than itself and more secret than every label which is an ancestor of it in the lattice. Currently, the lattice is explicitly defined as part of the \ifclib library. If a different secrecy label structure was needed, \ifclib would need to be modified. This limitation is purely an implementation burden and possible future work for \ifclib would be to allow the user to define the lattice themselves.

%\begin{figure*}
%    \centering
%    \caption{Example lattice structures labeled with A, B, and C base secrecy levels}
%    \includegraphics[width=0.7\linewidth]{lattice.png}
%    \label{fig:lattices}
%\end{figure*}

\iffalse
\subsubsection*{A Straightforward, but Inadequate, Approach for Allowing Legal Flows}

With secret values and labels as defined above, \ifclib could provide a variety of secrecy-preserving functions for applications to use. For example, \ifclib could provide a function \lstinline{sum} to add two secret integers and return the result as a secret integer, which would have the following definition:
\begin{lstlisting}
pub fn sum<Param1Label, Param2Label, ReturnLabel>(param1: Secret<i64, Param1Label>,
param2: Secret<i64, Param2Label>) -> Secret<i64,ReturnLabel>
  where ReturnLabel: MoreSecretThan<Param1Label> + MoreSecretThan<Param2Label> {
    return Secret::<i64, ReturnLabel>::new(param1.unwrap() + param2.unwrap());
}
\end{lstlisting}
The function \lstinline{sum} takes two secret, signed 64-bit integer values as input and returns a new secret value that is the sum of the two values.\footnote{While the code is specific to signed 64-bit integers for ease of presentation, it could be generalized to other numeric values.}
The return value is more secret than both of the inputs, which is ensured by the type constraint in the \lstinline{where} clause. Note that \lstinline{Secret::unwrap()},
%\mike{Should it just be unwrap()?} \ada{Yep, should only be unwrap_ref() if we're passing a reference to it}.
which returns \lstinline{Secret::value}, is a private method and cannot be executed directly from application code.

%\mike{Should the code snippets here pass references, to minimize differences with the example in the next subsection?}

Application code could call \lstinline{sum} to add two secret values and return a secret value as the following example shows:
\begin{lstlisting}
fn example(x: Secret<i64, Label_A>, y: Secret<i64, Label_B>) {
  let s: Secret<i64, Label_AB> = sum(x, y);
  println!("Sum is {:?}", s.declassify());
}
\end{lstlisting}
(The code is enclosed in a function to help concretize the example and to show that \lstinline{x} and \lstinline{y} are variables with labels $\{A\}$ and $\{B\}$, respectively, without having to be specific about where their values come from.) The application must specify the type for \lstinline{s} (i.e., \lstinline{Secret<i64, Label_AB>}) because the compiler cannot infer it uniquely (e.g., \lstinline{Label_ABC} would also be valid). The code will only compile if \lstinline{z} is more secret than both \lstinline{x} and \lstinline{y}, because of the \lstinline{where} clause in \lstinline{sum}'s definition.

Although this approach would allow the library to provide secrecy-preserving operations on built-in types (i.e., types in the Rust Standard Library)
for applications to use, it would not allow applications to define their own operations on custom secret types. Furthermore, the library would need to provide functions for every conceivable operation on every built-in type, and applications would need to build secret data structures entirely from \lstinline{Secret} instances of built-in types. Even more disqualifying is that this approach would not allow applications to express implicit flows.
\fi

\subsubsection*{Accessing Secrets in Secret Blocks}

To help restrict illegal flows and to limit the annotation burden,
\lstinline{Secret}-wrapped values may be accessed only inside of lexically scoped blocks of code called \emph{secret blocks}, as shown in \autoref{subfig:model:secret-blocks}.
% A secrect block is a DIFC principal that inherits its capabilities from its parent principal (the process, or another secret block in the case of nested blocks).
A secret block is a macro call
% \footnote{In Rust, every macro name ends with \lstinline{!}.}
that specifies a static label that restricts the values the block's body may read and write---the body
may only read values with lower or same secrecy, and may only write values with higher or same secrecy---prohibiting illegal explicit and implicit flows.
The block evaluates to a secret value with the the same label as the block.
% \maxt{What does it mean for a secrecy label to be ``uniform''? And how does that prevent flows from high to low secrecy?} \mike{Revised. How's that?}
% For example, the \lstinline{if}--\lstinline{else} in \autoref{lst:implicit flow} would need to be placed entirely in a secret block, whose label would need to be the same high-secrecy label as \lstinline{sec}.
% \mike{Maybe that's confusing.}
\autoref{fig:secure-cal} has two secret blocks (line~\ref{line:secure-cal:first-secret-block} and lines~\ref{line:secure-cal:begin-second-secret-block}--\ref{line:secure-cal:end-second-secret-block}), each with secrecy label $\{ a, b \}$.

% An application denotes a secret block with a macro call:\footnote{In Rust, every macro name ends with \lstinline{!}.}
% \begin{lstlisting}
% secret_block!($\mathit{label}$ { $\mathit{body}$ })
% \end{lstlisting}
% where $\mathit{label}$ is the type representing the block's secrecy label, and $\mathit{body}$ is application code that can operate on secret values.
% The block evaluates to a value with type \lstinline|Secret<T,$\mathit{label}$>| for any type \lstinline|T|.

The secret block's body can call \lstinline{unwrap_secret($e$)} on an expression $e$ with type \lstinline|Secret<Type,$\mathit{label}$>|, as long as $\mathit{label}$ is no more secret than the block's secrecy label.
\Ifclib also provides variants of \lstinline{unwrap_secret}
% (given in \autoref{subfig:model:unwrap})
that take and return references instead of values, called \lstinline{unwrap_secret_ref} and \lstinline{unwrap_secret_mut_ref}.
(\autoref{sec:motivating-example} explained immutable and mutable references briefly.)
% \maxt{I'm not sure we should assume our readers understand Rust references.} \mike{Addressed with back ref.}
\autoref{fig:secure-cal}'s second secret block (lines~\ref{line:secure-cal:begin-second-secret-block}--\ref{line:secure-cal:end-second-secret-block}) calls \lstinline{unwrap_secret} and its variants to access secret values with labels $\{ a \}$, $\{ b \}$, and $\{ a, b \}$---all of which are no more secret than the block's label $\{ a , b \}$.

Note that code \emph{outside} of a secret block cannot call \lstinline{unwrap_secret()} or its variants because they are \emph{not} defined functions---they are recognized only by \lstinline{secret_block!}, as described in \autoref{subsec:ensure-noninterference:restrict-secret-accesses}.
% Code cannot call the \lstinline{unsafe} function \lstinline{Secret::unwrap()} directly without using the \lstinline{unsafe} keyword; code that uses \lstinline{unsafe} forgoes \ifclib's guarantees (and Rust's safety guarantees) and must be audited (\autoref{sec:requirements-and-assumptions}).

A secret block calls \lstinline{wrap_secret($e$)} to create a new \lstinline{Secret} value with the same secrecy label as the block.
\autoref{fig:secure-cal}'s first secret block (line~\ref{line:secure-cal:first-secret-block}) evaluates to a secret value created by a call to \lstinline{wrap_secret}. The second secret block (lines~\ref{line:secure-cal:begin-second-secret-block}--\ref{line:secure-cal:end-second-secret-block}) has no return value.\footnote{Technically, the block's body's return type is the empty type, \lstinline{()}, which implements \lstinline{SecretTrait<L>}, as required by the secret block (\autoref{subfig:secret-block-expansion}).}

A secret block can call application-defined functions only if they have been annotated as \emph{side effect free} using a \ifclib-provided attribute macro, \lstinline{#[side_effect_free_attr]} (\autoref{subfig:model:side-effect-free-attr}). Side-effect-free functions in turn can only call other side-effect-free functions.

To use a custom type in secret blocks, an application must annotate the type as side effect free
by annotating it with \lstinline|#[derive(InvisibleSideEffectFree)]| (\autoref{subfig:model:derive-isef}).
% as \autoref{sec:application-defined-functions-invisible} describes in more

The running example in \autoref{fig:secure-cal} does not define any side-effect-free functions or types, but Figs.~\ref{fig:game_state_ds}--\ref{fig:bs_game_loop} in \autoref{subsec:eval-battleship} do.

\subsubsection*{Declassification}

\Ifclib allows high-secrecy values to flow to low-secrecy values through the \lstinline{declassify()} method and its variants (\autoref{subfig:model:declassify}), which return the unwrapped value.
Line~\ref{line:secure-cal:declassify} in \autoref{fig:secure-cal} declassifies the secret value of \lstinline{count}.
Declassification is a trusted operation and should be audited by developers.

% \mike{I don't think it's sound to allow declassification outside of a secret block, because of untrusted macros operating on the code. But declassification already requires auditing, so it's not so bad? Anyway we should discuss this when talking about macro expansion and maybe in Sections~\ref{subsec:security-guarantees-threats} and/or \ref{subsec:limitations} too.}
% \mike{Discussed in \autoref{subsec:security-guarantees-threats}, which is also where we talk about how macro expansion order works.}

\iffalse
\autoref{fig:Example_With_Secrets} shows example code that uses \ifclib. The example code compares two secret integers and prints a string that depends on the result of the comparison. It defines a secret block that unwraps two secret integers
% (\lstinline{Secret<i64, Label_A>} and \lstinline{Secret<i64, Label_B>})
and returns a secret string with label $\{a, b\}$ (\lstinline{Secret<&str, Label_AB>}).
The secret block in turn calls a side-effect-free function \lstinline{calc_diff}, that returns the difference of two ordinary (non-secret) integers. (Introducing a separate call to compute the difference is completely unnecessary, but it helps us illustrate the programming model.)
The example finally calls \lstinline{Secret::declassify()} on the resulting secret string,
% (using a declassification method that simply returns the unwrapped, lowest-secrecy value),
illustrating how the application can explicitly declassify and then pass a secret value to an untrusted (i.e., external) entity.

\begin{figure}
\begin{lstlisting}
fn example(x: Secret<i64, Label_A>, y: Secret<i64, Label_B>) {
  let compare = secret_block!(Label_AB {
    let diff = calc_diff(unwrap_secret(x), unwrap_secret(y));
    if diff > 0i64 { // 0i64 is the constant 0 of type i64
      wrap_secret("x wins!")
    } else {
      wrap_secret("x loses!")
    }
  });
  println!("Result: {:}", compare.declassify());
}

#[side_effect_free_attr]
fn calc_diff(op1: i64, op2: i64) -> i64 {
  op1 - op2
}
\end{lstlisting}
\caption{A simple example of application code that uses \ifclib.}
%\ada{To compile, these would need to use \lstinline{declassify_ref()} in the print statements. Update: Not a syntactic requirement. Println's would compile with \lstinline{declassify_consume()} (abbreviated here as \lstinline{declassify()} at the cost of, well, consuming the variable.}
\label{fig:Example_With_Secrets}
\end{figure}
\fi

\iffalse
\mike{Reviewer says the following:}
\begin{lstlisting}[breaklines=true,language=TeX]
I think there is a missed opportunity to simultaneously (a) improve the quality of the system description, and (b) hint at the bones of a correctness proof. I think Section 4 would be significantly improved if it had the following narrative structure:

> Imagine a statement `lhs = rhs` in a secret block. The core property that Cocoon seeks to enforce is that: if `lhs` is defined outside the block AND `rhs` is influenced by a secret at level `L1`, THEN `lhs` must be a secret with level `L2 >= L1`.
>
> Case on the possible values of `lhs` and `rhs`. If `lhs = *e` where `e: &mut T`, then `T` must have come from `unwrap_secret_mut_ref(..)` because...

As it stands, Section 4 still feels like all the details are little disconnected from each other.
\end{lstlisting}
\mike{I see the reviewer's point, except I think that starting with \lstinline{lhs = rhs} is starting with a particular type of expression (assignment expression), rather than being general by starting with expression $e$. Anyway, I made a few revisions along the lines suggested.}
\ada{Looks good to me.}
\fi

\subsection{How \IfcLib Ensures Noninterference}
\label{subsec:ensure-noninterference}

To ensure noninterference, \ifclib restricts the behavior of secret blocks.
Given a secret block
\lstinline|secret_block!($L$, $\mathit{body}$)|, \ifclib restricts the expression $\mathit{body}$ in the following ways:
\begin{itemize}

\item $\mathit{body}$ must conform to IFC rules, e.g., it cannot read a value with higher secrecy than the block.
\autoref{subsec:ensure-noninterference:restrict-secret-accesses} describes how \ifclib enforces IFC rules.

\item $\mathit{body}$ must not have side effects such as sending data to a socket or writing to a variable visible outside of the block.
% To enforce this rudimentary effect system, \ifclib leverages Rust's features in novel ways.
% We classify side effects as either \emph{syntactically visible or invisible} depending on whether they are visible to Rust's procedural macros, which operate on an untyped AST (\autoref{subsec:restrict-secret-accesses}).
\autoref{sec:application-defined-functions-visible} and \autoref{sec:application-defined-functions-invisible} describe how \ifclib enforces side effect freedom, essentially providing a rudimentary effect system.
% that captures whether operations could violate noninterference.

\end{itemize}
% This section describes how \ifclib restricts the flow of secret values in secret blocks, and \autoref{subsec:restrict-side-effects} describes how \ifclib restricts arbitrary side effects in secret blocks.
To enforce these restrictions, \ifclib leverages Rust \emph{procedural macros}, which are procedures that operate on compiled code at compile time. Type information is not available at macro expansion time, and so the procedural macro operates on an untyped abstract syntax tree (AST).
% \Ifclib's procedural macros, \lstinline|secret_block!| and \lstinline|#[side_effect_free_attr]|, operate on secret blocks and functions called transitively by secret blocks, respectively.
% (\lstinline|#[derive(InvisibleSide|\-\lstinline|EffectFree)]| is technically also a procedural macro, but it operates on \lstinline{struct} definitions instead of code blocks.)

Throughout \autoref{subsec:ensure-noninterference:restrict-secret-accesses}--\autoref{sec:application-defined-functions-invisible}, we describe and refer to Figs.~\ref{fig:outer-macro-transformations} and \ref{fig:transf-all-exprs}, which provide details of the procedural macros' transformations.

\subsubsection{Restricting the Flow of Secret Values}
\label{subsec:ensure-noninterference:restrict-secret-accesses}

\Ifclib enforces IFC rules on the flow of secret values using transformations by the \lstinline{secret_block!} procedural macro.

% Figs.~\ref{subfig:secret-block-expansion} and \ref{subfig:transf-unwrap-wrap} show transformations performed on secret blocks to mediate accesses to secret values.
\lstinline{secret_block!} expands \lstinline{unwrap_secret($e$)} to a call to a function that requires $e$'s label to be no more secret than the block's label $L$ (similarly for \lstinline{unwrap_secret_ref($e$)}).
For \lstinline{unwrap_secret_mut_ref($e$)}, the type constraint requires $e$'s label to be \emph{equal to} $L$ because mutable references allow both writing and reading the referent.
% %
\autoref{subfig:transf-unwrap-wrap} shows details of these transformations, by defining the function $\rewriteparam{e}{isExec}$ for when $e$ is an \lstinline{unwrap_secret} (or \lstinline{wrap_secret}) expression. In general, the text uses $\rewriteparam{e}{isExec}$ to define compile-time transformations performed by the macros on expression $e$ (the $\mathit{isExec}$ parameter is explained later).
% \lstinline|Secret::unwrap::<$L$>($e$)| (similarly for \lstinline{unwrap_secret_ref} and \lstinline{unwrap_secret_mut_ref}) where $L$ is the block's label.
% (\lstinline|Secret::unwrap()| is an \lstinline|unsafe| method and thus cannot be called directly without using the \lstinline|unsafe| keyword, forgoing \ifclib's guarantees.)
% The definition of \lstinline|Secret::unwrap($e$)| has a type constraint requiring that $e$'s label is no more secret than $L$.
% For \lstinline{unwrap_secret_mut_ref($e$)}, the type constraint requires $e$'s label to be \emph{equal to} $L$ because mutable references allow both writing and reading the referent.

\begin{figure}
\small
\lstset{basicstyle=\footnotesize\ttfamily}

\begin{lrbox}{\mylisting}
\begin{minipage}{\linewidth}
Application code:
\begin{lstlisting}
secret_block!($L$, $\mathit{body}$)
\end{lstlisting}

Macro-transformed code:
\begin{lstlisting}
if true {
  ::Cocoon::call_closure::<$L$,_,_>(|| -> _ {
    ::std::panic::catch_unwind(::std::panic::AssertUnwindSafe(|| {
      $\rewrite{\mathit{body}}$
    }))
  }.unwrap_or_default())
} else {
  ::Cocoon::call_closure::<$L$,_,_>(|| -> _ { $\rewritesimple{\mathit{body}}$ })
}
\end{lstlisting}

% \mike{Reviewer says: ``The illustration for code snippets can improve. It is not particularly receptive to non-Rust experts. For example [here] there is supposed to be a type constraint capturing that the block does not unwrap secrets greater than L; however it is not clear where the constraint annotation is.'' Hopefully the text is more clear now and doesn't give the impression that this figure is supposed to show \lstinline{unwrap_secret} type constraints.}

Definition of \lstinline{call_closure}:
\begin{lstlisting}
fn call_closure<L, F, R>(clos: F) -> R
  where F: FnOnce() -> R + VisibleSideEffectFree, R: SecretTrait<L>
  { clos() }
\end{lstlisting}
Note: \lstinline{SecretTrait<L>} is a trait implemented by \lstinline{Secret<T,L>} and tuples of \lstinline{Secret<T,L>}, allowing secret blocks to return multiple secret values.
\end{minipage}
\end{lrbox}

\subfloat[How the \lstinline|secret_block!| macro expands secret blocks. As the text explains, the macro generates both executed (\lstinline{if true}) and nonexecuted (\lstinline{else}) versions of the block's body.]{
  \usebox{\mylisting}
  \label{subfig:secret-block-expansion}
}\\[5em]

%\newsavebox{\mylisting}
\begin{lrbox}{\mylisting}
\begin{minipage}{\linewidth}
Application code:
\begin{lstlisting}
#[side_effect_free_attr]
fn $f$($\mathit{params}$) -> $r$ { $\mathit{body}$ }
\end{lstlisting}

Macro-transformed code:
\begin{lstlisting}
fn __$f$_secret_trampoline_unchecked($\mathit{params}$) -> $r$ { $\rewritesimple{\mathit{body}}$ }
fn __$f$_secret_trampoline_checked($\mathit{params}$) -> $r$ { $\rewrite{\mathit{body}}$ }
#[inline(always)]
unsafe fn $f$($\mathit{params}$) -> ::Cocoon::Vetted<$r$> {
    ::Cocoon::Vetted::<$r$>::wrap(__$f$_secret_trampoline_unchecked($\mathit{args}$))
}
\end{lstlisting}
\end{minipage}
\end{lrbox}

\subfloat[How the \lstinline|\#\[side_effect_free_attr\]| macro expands a side-effect-free function $f$. As the text explains, the macro generates both executed (\lstinline|__$f$_secret_|\allowbreak\lstinline|trampoline_unchecked|) and nonexecuted (\lstinline|__$f$_secret_trampoline_checked|) versions of $f$.]{
  \usebox{\mylisting}
  \label{subfig:side-effect-free-attr-expansion}
}\\[3em]

% \mike{Reviewer says: ``this figure references the $\tau$ operator which hasn't been introduced. And as far as I can tell, it doesn't get explicitly introduced anywhere in the prose.'' It's now introduced in the text and in this figure's caption.}

\caption{Outer transformations performed by \ifclib's procedural macros on secret blocks and side-effect-free functions.
$\rewriteparam{\mathit{expr}}{isExec}$ generates transformed code for expression $\mathit{expr}$ at macro expansion time, where $\mathit{isExec}$ is a boolean indicating whether the transformation is for the executed or nonexecuted version of generated code. \autoref{fig:transf-all-exprs} details how $\rewriteparam{\mathit{expr}}{isExec}$ is defined for various kinds of expressions.}
\label{fig:outer-macro-transformations}
\end{figure}

\begin{figure}
\footnotesize
\lstset{basicstyle=\footnotesize\ttfamily}

\newcolumntype{H}{>{\setbox0=\hbox\bgroup}c<{\egroup}@{}}
\newcolumntype{Z}{>{\setbox0=\hbox\bgroup}c<{\egroup}@{\hspace*{-\tabcolsep}}}
\newcommand{\mc}[2]{\multicolumn{#1}{c}{#2}}

%\newsavebox{\mylisting}
\begin{lrbox}{\mylisting}
%\begin{minipage}{\linewidth}
\begin{tabular}{l|lZ}

$\bm{\mathit{expr}}$ & \boldmath $\rewriteparam{\mathit{expr}}{isExec}$ & \textbf{Comment} \\\hline

\lstinline|unwrap_secret($e$)| &
\lstinline|{ let tmp = $\rewriteparam{e}{isExec}$; unsafe { Secret::unwrap::<$L$>(tmp) } }| &
Type constraint: $L \supseteq e$'s label \\\hline

\lstinline|unwrap_secret_ref($e$)| &
\lstinline|{ let tmp = $\rewriteparam{e}{isExec}$; unsafe { Secret::unwrap_ref::<$L$>(tmp) } }| &
Type constraint: $L \supseteq e$'s label \\\hline

\lstinline|unwrap_secret_mut_ref($e$)| &
\lstinline|{ let tmp = $\rewriteparam{e}{isExec}$; unsafe { Secret::unwrap_mut_ref::<$L$>(tmp) } }| &
Type constraint: $L = e$'s label  \\\hline

\lstinline|wrap_secret($e$)| &
\lstinline|{ let tmp = $\rewriteparam{e}{isExec}$; unsafe { Secret::new::<_,$L$>() } }| &

\end{tabular}
%\end{minipage}
\end{lrbox}

\subfloat[Transformations performed by \ifclib's procedural macros on secret blocks only (i.e., not on side-effect-free functions). $L$ is the secret block's label. \lstinline{Secret} is an abbreviation for \lstinline{::cocoon::Secret}.]{
  \usebox{\mylisting}
  \label{subfig:transf-unwrap-wrap}
}

%\newsavebox{\mylisting}
\begin{lrbox}{\mylisting}
%\begin{minipage}{\linewidth}
\begin{tabular}{p{1.2in}|p{4in}Z}

$\bm{\mathit{expr}}$ & \boldmath $\rewriteparam{\mathit{expr}}{\mathit{isExec}}$ & \textbf{Comment} \\\hline

\lstinline|callee($e_1$, $e_2$, $\dots$)| \linebreak if \lstinline|callee| is \emph{not} allowlisted &
\lstinline|let tmp1 = $\rewriteparam{e_1}{\mathit{isExec}}$; let tmp2 = $\rewriteparam{e_2}{\mathit{isExec}}$; $\dots$ |\allowbreak\lstinline|unsafe { callee(tmp1, tmp2, $\dots$) as ::cocoon::Vetted<_> .unwrap() }| &
If \lstinline|callee| is \emph{not} allowlisted \\\hline

\lstinline|callee($e_1$, $e_2$, $\dots$)| \linebreak if \lstinline|callee| is allowlisted &
\sbscheck\lstinline|(callee($\rewriteparam{e_1}{\mathit{isExec}}$, $\rewriteparam{e_2}{\mathit{isExec}}$, $\dots$))| &
If \lstinline|callee| is allowlisted \\

\end{tabular}
%\end{minipage}
\end{lrbox}

\subfloat[Transformations performed by \ifclib's procedural macros on calls.]{
  \usebox{\mylisting}
  \label{subfig:transf-calls}
}

%\newsavebox{\mylisting}
\begin{lrbox}{\mylisting}
%\begin{minipage}{\linewidth}
\begin{tabular}{p{0.88in}|p{3in}|p{1.18in}Z}

$\bm{\mathit{expr}}$ & \boldmath $\rewrite{\mathit{expr}}$ & \boldmath $\rewritesimple{\mathit{expr}}$ & \textbf{Comment} \\\hline

\lstinline|literal|~~~(constant) &
\lstinline|literal| &
\lstinline|literal| &
Constant
\\\hline

\lstinline|path|~~~(var.\ name) &
\lstinline|{ let tmp = &path; unsafe { |\sbscheckunsafe\lstinline|(tmp) } }| &
\lstinline|path| &
Variable name\\\hline

\lstinline|$e_1$ + $e_2$| &
\lstinline|::Cocoon::SafeAdd::safe_add|\allowbreak\lstinline|($\rewritenocheck{e_1}$, $\rewritenocheck{e_2}$)| &
\lstinline|$\rewritesimple{e_1}$ + $\rewritesimple{e_2}$| &
Overloadable operator \\\hline

\lstinline|$e_1$ && $e_2$| &
\lstinline|$\rewritenocheck{e_1}$ && $\rewritenocheck{e_2}$| &
\lstinline|$\rewritesimple{e_1}$ && $\rewritesimple{e_2}$| &
Non-overloadable operator \\\hline

\lstinline|$e_1$ = $e_2$| &
\lstinline|*check_not_mut_secret(&mut $\rewritenocheck{e_1}$) |\allowbreak\lstinline|= $\rewrite{e_2}$ | &
\lstinline|$\rewritesimple{e_1}$ = $\rewritesimple{e_2}$| &
Assignment\\\hline

\lstinline|&|$e$ &
\sbscheck\lstinline|(&(|\rewritenocheck{e}\lstinline|))| &
\lstinline|&|$\rewritesimple{e}$ &
Reference expression\\\hline

\lstinline|*|$e$ &
\lstinline|*(|\rewrite{e}\lstinline|)| &
\lstinline|*|$\rewritesimple{e}$ &
Dereference expression\\\hline

\lstinline|$e$.field| &
\lstinline|$\rewrite{e}$.field| &
\lstinline|$\rewritesimple{e}$.field| &
\sbscheck\lstinline|($\mathit{expr}$)| not needed \\\hline

\lstinline|if $e_1$ { $e_2$ } |\allowbreak\lstinline|else { $e_3$ } | &
\lstinline|if $\rewrite{e_1}$ { $\rewrite{e_2}$ } |\allowbreak\lstinline|else { $\rewrite{e_3}$ }| &
\lstinline|if $\rewritesimple{e_1}$ { $\rewritesimple{e_2}$ } |\allowbreak\lstinline|else { $\rewritesimple{e_3}$ }| &
Conditional

\end{tabular}
%\end{minipage}
\end{lrbox}

\subfloat[Transformations performed by \ifclib's procedural macros on secret blocks and side-effect-free functions.]{%
  \usebox{\mylisting}%
  \label{subfig:transf-other-exprs}%
}

\iffalse
%\newsavebox{\mylisting}
\begin{lrbox}{\mylisting}
%\begin{minipage}{\linewidth}
\begin{tabular}{l|p{4in}}
\textbf{Transformation function} & \textbf{Behavior} \\\hline
\rewrite{e} & Defined in (a) above \\\hline
\rewritenocheck{e} & Same as \rewrite{} except outer \sbscheck{} is omitted \\\hline
\rewritesimple{e} & Same as \rewrite{} but recursively omits all \sbscheck{} and overloaded operator replacement
\end{tabular}
%\end{minipage}
\end{lrbox}

\subfloat[Definitions of transformation functions.]{
  \usebox{\mylisting}
  \label{subfig:transf-defs}
}
\fi

\flushleft

\caption{Expression-level transformations performed by \ifclib's procedural macros on secret blocks and side-effect-free functions. \rewrite{\mathit{expr}} and \rewritesimple{\mathit{expr}} represent syntactic expansions for the nonexecuted and executed code paths, respectively. \sbscheck{} stands for ``check invisible side effect free.''}
\label{fig:transf-all-exprs}
\end{figure}

% \maxt{Does our reader understand Rust's private fields?} \mike{Addressed}

% Code \emph{outside} of a secret block cannot call \lstinline{unwrap_secret()} because the function does not exist (it is only recognized by \lstinline{secret_block!}). Code cannot call the \lstinline{unsafe} function \lstinline{Secret::unwrap()} directly without using the \lstinline{unsafe} keyword; code that uses \lstinline{unsafe} forgoes \ifclib's guarantees (and Rust's safety guarantees) and must be audited (\autoref{sec:requirements-and-assumptions}).
% \maxt{We use the term ``unsafe'' but do not explain it very well. Unsafe rust code is always unsafe. I want us to make this clear so that it doesn't seem odd to reviewers unfamiliar with Rust that unsafe forgoes Cocoon's guarantees.}
% \ada{are you looking for an in-depth explanation of unsafe or is adding a phrase to the sentence, sufficient (e.g., ``Code cannot call the unsafe function \lstinline{Secret::unwrap()} directly without using the unsafe keyword; code that uses unsafe forgoes Rust's guarantees, and thus Cocoon's, and must be audited (\autoref{sec:requirements-and-assumptions})?''}
% \mike{Seems like we'd want to make it clearer in \autoref{sec:requirements-and-assumptions}, but adding those words here seems fine too.}
% \mike{Addressed here and there}
% Application code cannot access \lstinline{Secret}'s value field directly because it is private.

To ensure that a secret block returns a \lstinline{Secret} with the same label as the block, the macro transformation of \lstinline|secret_block!($L$ { $\mathit{body}$ })| ensures that $\mathit{body}$ returns a \lstinline{Secret} value with label $L$.
\autoref{subfig:secret-block-expansion} shows how this works: The macros transform the secret block to a call to \lstinline{Cocoon::call_closure()}, which must have type \lstinline|Secret<T,$L$>| (or a tuple of secrets with label $L$).
Note that the macros generate both \emph{executed} and \emph{nonexecuted} versions of the code (hence the \lstinline|if true {...} else {...}| in \autoref{subfig:secret-block-expansion}) for reasons explained in \autoref{sec:application-defined-functions-invisible}.

\subsubsection{Restricting Syntactically Visible Side Effects}\label{sec:application-defined-functions-visible}

To ensure that a secret block's body has no side effects that could violate noninterference, \ifclib's procedural macros perform transformations on the body. Here we describe how the macros handle side effects that are \emph{syntactically visible} to the macros, and \autoref{sec:application-defined-functions-invisible} describes how the macros handle side effects that are \emph{syntactically invisible} to the macros.

% \Ifclib restricts two kinds of syntactically visible side effects: escaping mutations and calls to functions with side effects.

\paragraph{Restricting Mutation-Based Side Effects}

To ensure that a secret block cannot perform writes visible outside of the block---except for writes to mutable \lstinline{Secret} values, which are allowed---\ifclib leverages Rust's ownership and mutability constraints.
\Ifclib prevents a secret block from modifying non-\lstinline{Secret} variables in its surrounding environment by forbidding the secret block's body, which is implemented as a closure, from capturing non-\lstinline{Secret} variables by mutable reference.
\Ifclib introduces the trait \lstinline{VisibleSideEffectFree} (\autoref{fig:traits}) to denote that the block does not capture variables in its environment by mutable reference.
(In Rust, a closure implements a trait only if all the variables it captures implement that trait.)
\Ifclib leverages Rust's auto traits and negative implementation features to define \lstinline{VisibleSideEffectFree} so that every type automatically implements it except for types that (transitively) contain non-\lstinline{Secret} mutable references (\lstinline{&mut T} except for \lstinline{&mut Secret<_,_>}) or interior-mutable cells (types that implement \lstinline{UnsafeCell}, e.g., \lstinline{RefCell<T>}).\footnote{Attempting to construct a custom ``unsafe cell'' type would require \lstinline{unsafe} (and would be undefined behavior), thus forgoing Cocoon's guarantees.} As a result, a closure that implements \lstinline{VisibleSideEffectFree} cannot mutably capture non-\lstinline{Secret} values.
% \autoref{fig:traits} lists the \lstinline{VisibleSideEffectFree} trait and a logical representation of the types it implements.
% The exception to this rule is that secret blocks may mutably capture a secret value (i.e., \lstinline{&mut Secret}). Since no other variable, besides the secret value(s), were captured mutably, the secret block can only modify the secret value.
The only possible side effect of mutation is through \lstinline{Secret} values, which \ifclib mediates.

To prevent a secret block from assigning to a mutable \lstinline{Secret} variable without going through \lstinline{unwrap_secret_mut_ref} (potentially leaking a more-secret value through an implicit flow), \lstinline{secret_block!} transforms every assignment expression \lstinline{$e_1$ = $e_2$} by wrapping $e_1$ in a call to \ifclib function \lstinline{check_not_mut_secret()}, as \autoref{subfig:transf-other-exprs} shows. The call requires statically that $e_1$'s type is (transitively) not \lstinline{Secret}, by using Rust auto traits and negative implementations.

\begin{figure}
\small
\lstset{basicstyle=\footnotesize\ttfamily}

% \subfloat[Structs. \lstinline{Secret} is part of the programming model, and \lstinline{Vetted} is used internally by \ifclib.]{
% \begin{tabular}{l|l|l}
% \textbf{Type name} & \textbf{Description} & \textbf{Name in impl.} \\\hline
% \lstinline|Secret<T: SecretValueSafe, L: Label>| & Secret value with type \lstinline|T| and label \lstinline|L| & \lstinline|Secret<T,L>| \\\hline
% \lstinline|Vetted<T>| & Wrapper for return value of side-effect-free functions & \lstinline|InfoLeakFreeReturn<T>|
% \end{tabular}
% \label{subfig:structs}
% }

\begin{tabular}{@{}p{0.9in}|p{1.9in}|p{2.3in}@{}}
\textbf{Trait} & \textbf{Description} & \textbf{Definition} \\\hline
\lstinline|VisibleSide|\-\lstinline|EffectFree| & Limits mutably captured values of secret blocks & \raggedright $($\lstinline|Immutable|$ \, \lor \, ($\lstinline|&mut Secret<_,_>|$)) \,\, \land$ \lstinline|&InvisibleSideEffectFree| \tabularnewline\hline
\lstinline|SecretValueSafe| & Restricts \lstinline|T| in \lstinline|Secret<T,L>| to immutable, block-safe types & \raggedright \lstinline|Immutable| $\land$ \lstinline|InvisibleSideEffectFree| \tabularnewline\hline
\lstinline|Immutable| & Types without interior or regular mutability & $\neg($\lstinline|UnsafeCell<_>|)$ \; \land \; \neg ($\lstinline|&mut _|$)$ \\\hline
\lstinline|InvisibleSide|\-\lstinline|EffectFree| & Types that can be used in secret blocks & Implemented individually for built-in and application types
\end{tabular}

\flushleft

%\mike{I changed \lstinline{InteriorImmutable} to \lstinline{Immutable} and changed definitions around a bit.}

% \mike{In the implementation, \lstinline{VisibleSideEffectFree} is positively implemented by \lstinline{&InvisibleSideEffectFree} and negatively implemented by $\neg$ \lstinline{&InvisibleSideEffectFree}. Not sure that that really makes any difference, since the macro needs to insert SBS checking on path expressions anyway. In any case, \lstinline{VisibleSideEffectFree} should probably be renamed since it's not covering everything in \lstinline{InvisibleSideEffectFree}.} \ada{Like \lstinline{VisibleSideEffectFree}?}
% \mike{Yeah that could work; we'd need to explain even more clearly why it includes \lstinline{InvisibleSideEffectFree}. :) Or \lstinline{SecretBlockCaptureable}?}
%\mike{Renamed \lstinline{SideEffectFree} to \lstinline{VisibleSideEffectFree}}

% \subfloat[Macros. All of the macros are part of the programming model.]{
% \lstset{basicstyle=\footnotesize\ttfamily}
% \begin{tabular}{p{1.5in}|l|l}
% \textbf{Type name} & \textbf{Description} & \textbf{Name in impl.} \\\hline
% \lstinline|secret_block!| & Lexically scoped block that can unwrap secrets & \lstinline|secret_block!| \\\hline
% \lstinline|#[side_effect_free_attr]| & Annotates side-effect-free functions & \lstinline|#[info_leak_free]| \\\hline
% \lstinline|#[derive(InvisibleSide|\-\lstinline|EffectFree)]| & Annotates application \lstinline|struct| % definitions & \lstinline|#[derive(SecretBlockSafeDerive)]|
% \end{tabular}
% \label{subfig:macros}
% }

\caption{Traits used by \ifclib to enforce programming model restrictions to ensure side effect freedom.}
% Traits for limiting types that can be used in secret blocks and \lstinline{Secret} structs.
% All traits are used internally by \ifclib, but programmers may need to understand the traits to meet code requirements in side-effect-free contexts.

\label{fig:traits}
\end{figure}

\Ifclib must limit mutation in secret blocks, but not in functions called (transitively) by secret blocks: In Rust, functions (unlike closures) cannot mutate memory accessible outside of the function except through parameters to the function.
% for example, writes to global variables require \lstinline{unsafe} code. Since a side-effect-free function's parameters must come from another side-effect-free context, any mutations are ultimately confined to the secret blocks that (transitively) call side-effect-free functions.

\Ifclib must also ensure that values of type \lstinline{T} wrapped in a \lstinline{Secret<T, Label>} are not
interior mutable to avoid, for example, a secret block that reads high-secrecy data using \lstinline{unwrap_secret()} instead of \lstinline{unwrap_secret_mut()} to modify an interior-mutable value with low secrecy.
% In Rust, all interior-mutable types use an underlying structure that implements the \lstinline{UnsafeCell} trait.
\Ifclib disallows interior-mutable types in a \lstinline{Secret} by requiring that \lstinline{T} implement a \ifclib-defined trait \lstinline{SecretValueSafe} (\autoref{fig:traits}) that disallows mutable types.
%\Ifclib disallows this by requiring that \lstinline{T} implement a trait \lstinline{SecretValueSafe}, which is only implemented for types that implement the trait \lstinline{InteriorImmutable}. All types implement \lstinline{InteriorImmutable} unless they use the \lstinline{UnsafeCell} trait, which is a trait implemented by all types (notably \lstinline{RefCell<T>}) that use interior mutability.
% \autoref{fig:traits} shows the \lstinline{SecretValueSafe} trait.
% The figure shows that \lstinline{SecretValueSafe} is also only implemented for types that implement \lstinline{InvisibleSideEffectFree}, a trait implemented by types that are allowed to be used in side-effect-free contexts (\autoref{sec:application-defined-functions-invisible}).
%\mike{Reviewer found (an older version of) this paragraph confusing.}

\paragraph{Prohibiting Calls to Functions with Side Effects}
\label{subsec:prohibiting-calls-with-side-effects}

To ensure side effect freedom, \ifclib must ensure that every function called (transitively) by secret blocks is side effect free.
Applications can annotate side-effect-free functions with the \lstinline{#[side_effect_free_attr]} macro (introduced in \autoref{subsec:programming-model} and \autoref{subfig:model:side-effect-free-attr}), which causes the function's body to be transformed to ensure side effect freedom, as \autoref{subfig:side-effect-free-attr-expansion} shows.
However, it is not straightfowrard to ensure that every function called from a \emph{side-effect-free context} (i.e., a secret block or side-effect-free function) is side effect free, because type information is not available at macro expansion time.

To address this challenge, the procedural macro \lstinline{#[side_effect_free_attr]} transforms the return type of the annotated function from \lstinline{R} to \lstinline{Vetted<R>}, a \ifclib-provided type. \Ifclib's procedural macros also rewrite all function calls within side-effect-free contexts from \lstinline{f(...)} to \lstinline{(f(...) as Vetted<R>).value}, mandating that \lstinline{f} return \lstinline{Vetted<R>}. As a result, every callee function must have been marked side effect free.
% or the application will not compile.
\autoref{subfig:transf-calls} shows how the macros transform calls in side-effect-free contexts.

Note that unverified functions cannot get around this requirement by directly returning \lstinline{Vetted<R>}.
First, a function cannot simply construct a \lstinline{Vetted} value and return that function's return value because the constructor of \lstinline{Vetted} is marked as \lstinline{unsafe} to prevent unverified code from creating \lstinline{Vetted} return values.
Second, a function cannot call another function that has been transformed by the \lstinline{#[side_effect_free_attr]} macro to obtain a \lstinline{Vetted} value, because the macro marks each function it generates as \lstinline{unsafe} to disallow calling the vetted function directly (without using \lstinline{unsafe}).

%\ada{Double check that the following does(n't) apply for secret block.} \mike{Actually, isn't this paragraph really about \lstinline{#[side_effect_free_attr}, not \lstinline{side_effect_free_closure!}? Secret blocks only impact \lstinline{side_effect_free_closure!}, not \lstinline{#[side_effect_free_attr}.}
Since the \lstinline{#[side_effect_free_attr]} macro generates a function marked \lstinline{unsafe},
the untrusted function body could contain expressions that are only permitted in unsafe Rust, although the original function was not marked \lstinline{unsafe}.
% the macro takes precautions to preserve Rust's safety guarantees.
The \lstinline{#[side_effect_free_attr]} macro prevents this by generating a second function that contains the function body. This second function omits the \lstinline{unsafe} modifier (assuming the original function did the same), thus preventing safety violations. \autoref{subfig:side-effect-free-attr-expansion} shows how the the macro transforms an annotated function.

In addition to allowing calls to functions marked side effect free in side-effect-free contexts, the procedural macro allows calls to Rust Standard Library functions that the \ifclib developers (i.e., we) have ``allowlisted'' as being side effect free. \Ifclib's procedural macro includes a list of functions that we have determined to be side effect free.
Because the procedural macro operates on tokens and has no type information, calls to allowlisted library functions from side-effect-free contexts must be fully qualified (e.g., \lstinline{::std::string::String::len()}) to prohibit calls to same-named functions with side effects.
% A future implementation could potentially eliminate this requirement (\autoref{subsec:impl-limitations}).
Allowlisting of library functions adds to the trusted codebase and could be fragile (\autoref{subsec:security-guarantees-threats}).

Thus, applications can mark any function as side effect free, and the compiler will check that the function is side effect free and that side-effect-free contexts call only side-effect-free functions.
% by virtue of all of its callees being side effect free
% (and generate a compiler error if not).
In order for a secret block to have a side effect, e.g., by sending data to a socket, it must (transitively) call a function \emph{not} marked side effect free, which cannot pass the procedural macros' checks.

\subsubsection{Restricting Syntactically Invisible Side Effects}\label{sec:application-defined-functions-invisible}

So far we have described how \ifclib prohibits side effects that are visible to procedural macros. However, Rust provides a few syntactically \emph{invisible} ways to call functions---through overloaded operators, deref coercion, and custom destructors---which are impossible for procedural macros to detect.

To prohibit syntactically invisible side effects in side-effect-free contexts,
\Ifclib's procedural macros transform every expression in a way that ensures that it cannot compile if it has a syntactically invisible side effect. The procedural macros actually generate \emph{two versions} of code in side-effect-free contexts: one that executes and one that does not. In the \emph{executed} version, overloadable operators and expressions are \emph{not} transformed by the procedural macro (but \lstinline{unwrap_secret} and \lstinline{wrap_secret} calls are still expanded). This version is the code that actually executes. In the \emph{nonexecuted} version, the procedural macro \emph{does} transform operators and expressions as described below, but the generated code is unreachable. As such, the expanded version causes compiler errors if \ifclib's requirements are violated, but it never executes. \autoref{fig:outer-macro-transformations} shows the transformations performed on secret blocks and side-effect-free functions to generate the executed and nonexecuted versions. The executed code and nonexecuted code are generated by \rewriteparam{\mathit{expr}}{isExec}, which \autoref{fig:transf-all-exprs} provides details of.

\Ifclib generates executed and nonexecuted versions for two reasons.
First, transforming a \emph{path expression} (i.e., use of a variable) requires violating borrowing rules: The transformation inserts a call to \sbscheckunsafe, which needs to take a reference to the path but return the expression, as \autoref{subfig:transf-other-exprs}'s \lstinline{path} row shows.
It is generally unsafe to allow this generated code to execute; generating it in the nonexecuted version means it only affects whether the code compiles.
Second, avoiding most transformations in the executed version may reduce compile- and run-time costs of the transformations.

\paragraph{Restricting Overloaded Operators}
\label{subsec:restricting-overloaded-operators}

Rust allows applications to overload certain operators (e.g., \lstinline{+} and \lstinline{<}) by defining custom functions for their behavior. For example, if an application overloads the \lstinline{+} operator for certain types, the expression \lstinline{a + b} might no longer be side effect free, depending on the types of \lstinline{a} and \lstinline{b}.
% as the newly overloaded add function could send \lstinline{x} and \lstinline{y} to the screen before returning the sum.
% As a result, \ifclib cannot assume that overloadable operators are safe to use.
\Ifclib's procedural macro cannot detect whether \lstinline{a + b} invokes an overloaded operator because it operates on a parse tree without type information.
% Unlike allowing calls to side effect free functions, \ifclib cannot allowlist the Rust Standard Library operator functions because once the function has been overloaded, there is no way to recover the original Rust Standard Library function body.
To address this problem, \ifclib's procedural macro replaces all calls to \emph{overloadable} operators
% \footnote{Defined by Rust's \lstinline{std::ops} crate.}
with a call that implements the default operator behavior.

Specifically, \ifclib provides a series of internal traits (e.g., \lstinline{cocoon::SafeAdd} to replace \lstinline{std::ops::}\allowbreak\lstinline{Add}) that are essentially copies of the standard Rust traits for each overloadable operator.
% These traits are unsafe so as to disallow applications from implementing these traits on application or standard Rust types.
\Ifclib implements these traits for all applicable standard Rust types.
% Since Rust disallows implementing an external trait on an external type, applications cannot overload the implementation of the new safe traits on the standard Rust types. \mike{Seems redundant since the trait is unsafe.}
% \Ifclib's procedural macro replaces any use of an overloadable operator with a fully qualified call to the operator method in \ifclib---thus bypassing any overloading attempted by the application. Continuing with our addition example,
\Ifclib's procedural macros replace \lstinline{$e_1$ + $e_2$} with \lstinline{::Cocoon::SafeAdd::safe_add($e_1$, $e_2$)} in side-effect-free contexts. In this way, \ifclib allows the use of overloadable operators only on standard Rust types, for which the behavior is the original, Rust-standard side-effect-free behavior. Any use of these operators on application-defined types within a side-effect-free context
% (where we cannot ensure the operator has not been overloaded)
results in a compilation error. \autoref{subfig:transf-other-exprs} shows this transformation for \lstinline{$e_1$ + $e_2$}; the transformation is analogous for other overloadable operators.

Code cannot circumvent this restriction by overloading operators on built-in Rust types (e.g., numeric types): Overloadable operators are defined by built-in Rust traits, and Rust does not permit a crate to implement a trait for a type unless the crate defines the trait or the type (or both).

% It may be possible for \ifclib to support overloaded operators in side-effect-free contexts by extending the checking of side-effect-free functions,
% % (\autoref{sec:application-defined-functions-visible}),
% but our design and prototype implementation do not provide such support.

\paragraph{Restricting Deref Coercion and Custom Destructors}

One of Rust's features is that, in attempting to find a match for a function or struct signature, the Rust compiler performs \emph{deref coercion}, automatically inserting dereference calls until it finds a match for the signature or cannot dereference any further. For example, the compiler automatically converts \lstinline{s.area()} to \lstinline{s.deref().area()} if \lstinline{s} has type \lstinline{&Shape} and method \lstinline{Shape::area()} exists. Likewise, the compiler converts \lstinline{b.area()} to \lstinline{b.deref().deref().area()} if \lstinline{b} has type \lstinline{&Box<Shape>} and method \lstinline{Shape::area()} exists. This situation presents a challenge because a call to \lstinline{s.area()} can implicitly invoke \lstinline{s.deref()} or \lstinline{s.deref().deref()} in a way that is invisible to \ifclib's procedural macro. While \lstinline{Deref::deref()} and \lstinline{DerefMut::deref_mut()} have no side effects for built-in types, an application
% may implement \lstinline{Deref} or \lstinline{DerefMut} for an application type and
could define a \lstinline{deref()} or \lstinline{deref_mut()} method that has side effects.

Another problematic Rust feature is that applications can define custom destructor behavior. The compiler inserts a \emph{drop} (i.e., deallocation) of a value in the code where the value dies. The application can provide custom behavior on drop by implementing the \lstinline{Drop} trait on an application type and providing an implementation of \lstinline{Drop::drop()}---which could have a syntactically invisible side effect.
% if the implementation has side effects. A side-effect-free context could thus have a side effect if it drops an instance of the application type.
% The procedural macro cannot detect potential calls to a custom \lstinline{Drop::drop()} because the procedural macro executes before the compiler inserts drops.
% In this way, a type could overload Drop and be instantiated inside a \lstinline{side-effect-free-block} without explicitly leaking anything inside that block. However, when the instantiation was destroyed, it might print information from inside that \lstinline{side-effect-free-block} to the screen, without being detected by \ifclib. As such, only those types which don't implement the Drop trait, or which implement the Drop trait in such a way that we can be sure no information flow-related side effects occur, can be safely used inside a \lstinline{side-effect-free-block}.

\ignore{
Not only are deref coercion and destructor invocation invisible to \ifclib's procedural macros, but they can happen on values captured by \emph{and created within} side-effect-free contexts, so limiting the captures of secret block closures will not suffice.
}

To handle these operations,
the procedural macros insert code to ensure that \emph{every value used by or created in} a side-effect-free context does not provide a custom implementation of \lstinline{Deref}, \lstinline{DerefMut}, or \lstinline{Drop}.
% %
\Ifclib provides a trait \lstinline{InvisibleSideEffectFree} that can be implemented by any type that (1) does not provide a custom implementation of \lstinline{Deref::deref()}, \lstinline{DerefMut::deref_mut()}, or \lstinline{Drop::drop()} and (2) contains only \lstinline{InvisibleSideEffectFree} values (\autoref{fig:traits}).
\Ifclib implements \lstinline{InvisibleSideEffectFree} for conforming Standard Rust types. For example, \ifclib implements \lstinline{InvisibleSideEffectFree} for \lstinline{i64}, for \lstinline{Box<T: InvisibleSideEffect}\-\lstinline{Free>}, and for many other types. Conforming application types do not automatically implement \lstinline{InvisibleSideEffectFree}, but an application can annotate a type definition with \lstinline{#[derive(InvisibleSide}\allowbreak\lstinline{EffectFree)]} (\autoref{subfig:model:derive-isef}), which invokes a procedural macro that transforms the type's definition to ensure that (1) it does not implement \lstinline{Deref}, \lstinline{DerefMut}, or \lstinline{Drop} (by generating negative implementations of these traits) and (2) all of the type's contained values have types that implement \lstinline{InvisibleSideEffectFree} (by generating type constraints on the contained values).

%\sout{To ensure that a secret block captures only \lstinline{InvisibleSideEffectFree} values, \ifclib restricts the definition of \lstinline{VisibleSideEffectFree} (\autoref{sec:application-defined-functions-visible}) to \lstinline{InvisibleSideEffectFree} types, i.e., a type is \lstinline{VisibleSideEffectFree} only if it is \lstinline{InvisibleSideEffectFree} (in addition to the restrictions in \autoref{sec:application-defined-functions-visible}).}
%\mike{Not true. That doesn't work because of auto trait weirdness.}
% Similarly, to check values wrapped in \lstinline{Secret},
% \lstinline{SecretValueSafe} (\autoref{sec:application-defined-functions-visible}) is restricted to \lstinline{InvisibleSideEffectFree} types (in addition to the restrictions in \autoref{sec:application-defined-functions-visible}).

To ensure that side-effect-free contexts only use and create \lstinline{InvisibleSideEffectFree} values, \ifclib's procedural macro in general transforms every expression $e$ to \sbscheck\lstinline|($e$)| as Figs.~\ref{subfig:transf-calls} and \ref{subfig:transf-other-exprs} show.
% \Vincent{not actually every expression, there's at least a few that don't have this check.}
% to a call to an internal function \sbscheck\lstinline|()|.
The expression \sbscheck\lstinline|($e$)| simply returns the value $e$ evaluates to, but $e$'s type must implement \lstinline{InvisibleSideEffectFree} to compile.
% ensuring that side-effect-free contexts can only create values that are \lstinline{InvisibleSideEffectFree}.

Not every kind of expression needs to be wrapped in \sbscheck\lstinline|()|, as \autoref{subfig:transf-other-exprs} shows.
For example, the field expression \lstinline|$e$.field| does not need to be wrapped because it is \lstinline{InvisibleSideEffectFree} as long as $e$ is \lstinline{InvisibleSide}\-\lstinline{EffectFree} ($e$ still needs to be wrapped in \sbscheck\lstinline|()| in general). % \autoref{subfig:transf-other-exprs} shows the transformation for \lstinline|$e$.field|.
% \mike{Do we explain why only some kinds of expression in \autoref{subfig:transf-other-exprs} get wrapped in \sbscheck? (This is a separate issue from how the implementation avoids even some of the checks in the figure.)}
% \mike{Added}

Note that secret blocks are not prevented from \emph{capturing} non-\lstinline{InvisibleSideEffectFree} types.\footnote{Specifically, \lstinline{VisibleSideEffectFree}, which defines types that can be captured by secret blocks, does \emph{not} negatively implement all \lstinline{InvisibleSideEffectFree} types (because this does not seem possible due to limitations of auto traits and negative implementations). Instead, \lstinline{VisibleSideEffectFree} excludes implementation of all \lstinline{&T} where \lstinline{T} does \emph{not} implement \lstinline{Invisible}\-\lstinline{SideEffectFree} (\autoref{fig:traits}), which disallows captures of non-\lstinline{InvisibleSideEffectFree} variables by reference, but not by value.} However, it is sufficient for macro transformations to prevent secret blocks from \emph{creating} or \emph{using} such types.
% \Ifclib instead relies on macro transformations to disallow uses of non-\lstinline{InvisibleSideEffectFree} types in secret blocks.

\paragraph{Handling Exceptional Flow}
\label{subsec:handling-panics}

To avoid illegal implicit flows, a secret block must have a single static exit.
Note that even if the block's body executes \lstinline{return}, \lstinline{continue}, or \lstinline{break},
control always continues immediately after the block,
since a secret block's body is wrapped in a closure (\autoref{subfig:secret-block-expansion}).

A remaining problem are \emph{panics}, which Rust generates for unexpected or unrecoverable errors. By default, a panic aborts the execution, creating a termination channel if a secret block (or code it calls) panics. Worse yet, the application might \emph{catch} a panic outside of a secret block, creating an implicit flow. To avoid this, every secret block catches any panics within it (using \lstinline{std::panic::catch_unwind}). The secret block handles a caught panic by returning a \lstinline{Secret}-wrapped default value (\lstinline{Secret}-wrapped types must implement \lstinline{std::default::Default}) and continuing execution after the block normally, as \autoref{subfig:secret-block-expansion} shows.

\subsubsection{Example Transformation}

\begin{figure}
\raggedright\textsf{\small Expansion of \colorbox{lightgray}{\lstinline|secret_block!(lat::AB \{ wrap_secret(0) \})|}:}
\begin{lstlisting}
if true {
    (*@/* Expansion by \colorbox{lightgray}{$\rewritesimple{e}$} */@*)
    ::cocoon::call_closure::<lat::AB, _, _>(
        (|| -> _ {
            let result = ::std::panic::catch_unwind(::std::panic::AssertUnwindSafe(|| {
                (*@/* Expansion of \colorbox{lightgray}{wrap\_secret(0)} */@*)
                let tmp = 0; unsafe { ::cocoon::Secret::<_, lat::AB>::new(tmp) }
            })).unwrap_or_default();
            result
        }))
} else {
    (*@/* Expansion by \colorbox{lightgray}{$\rewrite{e}$} */@*)
    ::cocoon::call_closure::<lat::AB, _, _>(
        (|| -> _ {
          (*@/* Expansion of \colorbox{lightgray}{wrap\_secret(0)} */@*)
          unsafe { ::cocoon::Secret::<_, lat::AB>::new(0) }
        }))
}
\end{lstlisting}
\caption{The secret block at line~\ref{line:secure-cal:first-secret-block} of \autoref{fig:secure-cal},
after expansion by \ifclib's procedural macro \lstinline{secret_block!}.}
\label{fig:expanded-cal-first}
\end{figure}

\begin{figure}
\begin{lstlisting}
if true {
    (*@/* Expansion by \colorbox{lightgray}{$\rewritesimple{e}$} */@*)
    ::cocoon::call_closure::<lat::AB, _, _>(
        (|| -> _ {
            let result = ::std::panic::catch_unwind(::std::panic::AssertUnwindSafe(|| {
                if (*@/* Expansion of \colorbox{lightgray}{unwrap\_secret(available)} */@*)
                   { let tmp = available;
                   unsafe { ::cocoon::Secret::unwrap::<lat::AB>(tmp) } } &&
                   (*@/* Expansion of \colorbox{lightgray}{unwrap\_secret\_ref(::std::option::Option::unwrap(}@*)
                   (*@\colorbox{lightgray}{::std::collections::HashMap::get(\&bob\_cal, \&day)))} */@*)
                   *{ let tmp = ::std::option::Option::unwrap(
                                  ::std::collections::HashMap::get(&bob_cal, &day));
                      unsafe { ::cocoon::Secret::unwrap_ref::<lat::AB>(tmp) } }
                {
                  (*@/* Expansion of \colorbox{lightgray}{*unwrap\_secret\_mut\_ref(\&mut count) += 1;} */@*)
                  *{ let tmp = &mut count;
                     unsafe { ::cocoon::Secret::unwrap_mut_ref::<lat::AB>(tmp) } } += 1;
                }
            }))
            .unwrap_or_default();
            result
        })
    )
} else {
    (*@/* Expansion by \colorbox{lightgray}{$\rewrite{e}$} */@*)
    ::cocoon::call_closure::<lat::AB, _, _>(
      || -> _ {
          if (*@/* Expansion of \colorbox{lightgray}{unwrap\_secret(available)} */@*)
             unsafe {
               ::cocoon::Secret::unwrap::<lat::AB>(
                 { let tmp = &(available); unsafe { ::cocoon::check_ISEF_unsafe(tmp) } }) } &&
             (*@/* Expansion of \colorbox{lightgray}{unwrap\_secret\_ref(::std::option::Option::unwrap(}@*)
             *unsafe {
              ::cocoon::Secret::unwrap_ref::<lat::AB>({
                ::cocoon::check_ISEF(::std::option::Option::unwrap({
                  ::cocoon::check_ISEF(::std::collections::HashMap::get(
                      { ::cocoon::check_ISEF_ref(&bob_cal) },
                      { ::cocoon::check_ISEF_ref(&day) })) })) }) }
          {
            (*@/* Expansion of \colorbox{lightgray}{*unwrap\_secret\_mut\_ref(\&mut count) += 1;} */@*)
            ::cocoon::SafeAddAssign::safe_add_assign(
                &mut *(unsafe { ::cocoon::Secret::unwrap_mut_ref::<lat::AB>(
                                  { ::cocoon::check_ISEF((&mut count)) })
                }), 1);
          }
      })
}
\end{lstlisting}
\caption{The secret block at lines~\ref{line:secure-cal:begin-second-secret-block}--\ref{line:secure-cal:end-second-secret-block} of \autoref{fig:secure-cal},
after expansion by \ifclib's procedural macro \lstinline{secret_block!}.}
\label{fig:expanded-cal-second}
\end{figure}

Figs.~\ref{fig:expanded-cal-first} and \ref{fig:expanded-cal-second} show how the \lstinline|secret_block!| procedural macro transforms \autoref{fig:secure-cal}'s first and second secret blocks, respectively.

% \bigskip
% \noindent
% Our submitted supplementary material shows in detail how \ifclib's macros transform \autoref{fig:Example_With_Secrets}'s code to ensure side effect freedom. The macro-transformed code is not intended to be readable by developers, but serves as a concrete example of the result of the macro transformations described in Sections~\ref{sec:application-defined-functions-visible}--\ref{sec:secret-blocks}.

% \mike{It would be nice to have example(s) for Sections~\ref{sec:application-defined-functions-visible}--\ref{sec:secret-blocks}}.
% \ada{Added comment at beginning of 4.2 pointing out the appendix example. Seems redundant to keep referring to the example in each subsection. }

\iffalse
\paragraph{Allowing calls to side effect free functions using secret values}\ada{Do we still need the following paragraph - should we update it to instead discuss details of how the unwrap\_secret() and wrap\_secret() functions are correct? This would resolve my above comments.}
\mike{This part should be removed. I think discussion of secret blocks should go elsewhere---where exactly is it now?} \ada{I did not add any additional text regarding secret block - I just tried to replace any mentions of secret closure with secret block.}

Application code passes secret values to side-effect-free functions---which take ordinary (non-secret) values as input---by using one of the aforementioned \lstinline{apply} functions. Each \lstinline{apply} function takes secret values and a side-effect-free closure as parameters, unwraps the secrets, calls the closure on the unwrapped values, and returns the closure's return value wrapped as a secret. For example, \ifclib uses the following definition of \lstinline{apply_binary_ref}, which takes two secret parameters and returns a secret value:
\medskip\\
\begin{minipage}{\linewidth}
\begin{lstlisting}
pub fn apply_binary_ref<'a, F, P1T : IntImm, P1L, P2T : IntImm, P2L, RT : IntImm, RL>
       (f: SideEffectFreeClosure<F, RT, (&'a P1T, &'a P2T)>,
        param1: &'a Secret<P1T,P1L>, param2: &'a Secret<P2T,P2L>) -> Secret<RT,RL>
  where F: VisibleSideEffectFree + FnOnce((&'a P1T, &'a P2T)) -> RT,
        RL: MoreSecretThan<P1L> + MoreSecretThan<P2L> {
    let retval = f((param1.unwrap_ref(), param2.unwrap_ref()));
    Secret::<RT,RL>::new(retval)
}
\end{lstlisting}
\end{minipage}
\smallskip\\
The function's type constraints ensure that \lstinline{f} is a closure created with \ifclib's \lstinline{side_effect_free_closure!} macro: Only functions created with \lstinline{side_effect_free_closure!} have the type \lstinline{SideEffectFreeClosure}.
%\mike{Why is \lstinline{F: VisibleSideEffectFree} (previously called \lstinline{InfoLeakFree}) needed? I don't think the text mentions \lstinline{VisibleSideEffectFree}?}
%\maxwell{I tried to address this by expanding the content in \emph{Preventing writes.} Take a look, and see if that addresses the issue.}
The type constraints also ensure that the return value is more secret than both of the parameter types. In \lstinline{apply_binary_ref}, the parameters are passed as borrowed immutable references, which is convenient for applications that want to keep using the values (otherwise the values would be consumed by the function and could not be used subsequently by the application).
The body of \lstinline{apply_binary_ref} unwraps the parameters' secret values using the \lstinline{Secret::unwrap_ref()} method that is accessible from inside \ifclib. Then it calls the side-effect-free closure \lstinline{f} with the unwrapped values, and it returns the closure's return value wrapped as a secret value.
\lstinline{IntImm} is an abbreviation for \lstinline{InteriorImmutable}.
\fi

\iffalse
\mike{I think we don't need any of this:}
\lstinline{Secret} has the following definition:
\begin{lstlisting}
pub struct Secret<Type, SecrecyLabel> where Type: SecretValueSafe {
    value: Type,
    _unused: PhantomData<SecrecyLabel>
}
\end{lstlisting}
First, note that ordinary encapsulation prevents application code from directly accessing \lstinline{value}, which is private.
% \Ifclib provides facilities for applications to perform secure operations on a \lstinline{Secret}'s \lstinline{value} field (discussed later).
Second, \ifclib requires the secret value's type to implement a trait, \lstinline{SecretValueSafe}, which implies a few restrictions that most types already meet.
% , such as being interior immutable and having no implementations of traits that allow side effects.
These restrictions are discussed in detail in \autoref{sec:application-defined-functions-visible} and \autoref{sec:secret-blocks}.
%Most types in the Rust Standard Library
%\mike{Incomplete sentence. Remove?} \ada{Removed. Reviewers didn't like ``most'' but I don't have a way of quantifying how many types use UnsafeCell. }
%\ada{Per reviewer comment: clearly state which types are not included and/or a percentage} are already \lstinline{SecretValueSafe}, and applications can mark custom types as \lstinline{SecretValueSafe} using a single \lstinline{#[derive]} annotation.
%$\ada{Update section reference.} \ada{say what secret value safe means, or reference a later section.}
% Most types in the standard Rust library  be \emph{interior immutable}---which applies to most Rust types, but does not apply to a few types (notably the \lstinline{RefCell} type). Without this restriction, application code could erroneously be allowed to store more-secret values in less-secret, interior-mutable variables.
% \lstinline{InteriorImmutable} is a \ifclib-defined Rust trait that all types implement except for interior mutable types, by using Rust's negative trait implementations (\autoref{subsec:impl-details}).

Note that \lstinline{PhantomData} is a zero-sized type provided by Rust to ensure the compiler parameterizes \lstinline{Secret} on the generic type \lstinline{SecrecyLabel}.
% Thus although \lstinline{Secret} conceptually ``wraps'' a value with a secrecy label,
Thus a \lstinline{Secret<Type, SecrecyLabel>} instance has the same run-time representation as a \lstinline{Type} instance.
\fi

\subsection{Implementation Details}
\label{subsec:impl-details}

Our prototype implementation of \ifclib is divided into two Rust crates---one that defines the procedural macros
% \lstinline{secret_block!}, \lstinline{#[side_effect_free_attr]}, and \lstinline{#[derive(InvisibleSideEffectFree)]};
and another that provides all other \ifclib functionality---since
% defines the \lstinline{Secret} type, associated operations, the secrecy label lattice, and the implementations for \lstinline{VisibleSideEffectFree} and other traits; and another that
procedural macros must be defined in their own crate.
% \mike{I think \ifclib is not a package because \ifclib contains two library crates, and by definition a package can contain at most one library crate.}
% %
The implementation depends on a \emph{nightly} version of Rust
(\nightlyversion)
% from \nightlydate)
since it uses a few Rust features that are not yet stable: auto traits, negative traits, function traits, and unboxed closures.
% %
The implementation's source code is publicly available.\footnote{\url{https://github.com/PLaSSticity/Cocoon-implementation}}

\subsection{Security Guarantees and Threats}
\label{subsec:security-guarantees-threats}

\Ifclib's general guarantee (which we have not proved) is termination-insensitive noninterference
% \mike{Reviewer says that statements like ``''\ifclib aims to provide termination-insensitive noninterference'' ``are not insightful without an actual/formal statement.'' Revised.}
in the absence of declassify operations and unsafe Rust code (\autoref{subsec:goals-soundness}). Here we discuss a few caveats of \ifclib's guarantees.

\subsubsection*{Trusted Codebase}

Since we have not proved that allowlisted Rust Standard Library functions are side effect free,
allowlisted functions must be trusted to be side effect free (just like \ifclib's codebase must be trusted).
Identifying library functions as side effect free can be tricky. For example, \lstinline{Vec::sort} should not be allowlisted because it calls an implementation of \lstinline{Ord::cmp} passed as a parameter, which could be an application-defined function with side effects. As another, hypothetical example, a function with a side-effect-free specification might have side effects internally (and perhaps only in a future implementation of the library).
% In future work, conservative static analysis may able to check side effect freedom of library functions automatically.

% \mike{We wrote the following in the rebuttal:}
% \begin{lstlisting}[breaklines=true,language=TeX]
% We agree that an application developer's "only recourse is to either petition the system developers, or reimplement `Vec::sort` from % the whitelisted methods." Our implementation currently only allowlists functions encountered in our case studies, but ideally Cocoon % should be released with as many side-effect-free functions allowlisted as possible.
%
% Incidentally, `Vec::sort` actually should *not* be allowlisted because it calls an implementation of `Ord::cmp`, which could be an application-defined function with side effects. For similar reasons, we need to (and will) remove a few allowlisted functions from our implementation: `<[T]>::sort`, `Iterator::by_ref`, `Iterator::next`, and `Iterator::take` (the last three functions are on a
% trait, but only calls to functions on a struct should be allowlisted).
% \end{lstlisting}
% \mike{TODO: Incorporate into text.}

In general, every component of an application (whether part of the application's crate or an external crate) must be ported to \ifclib, or it must be explicitly trusted by declassifying secret data before sending it to the component.
% Only allowlisted functions and functions \ifclib checked to be side effect free are available for use inside a secret block.
Porting components to \ifclib provides an incremental mechanism for reducing the trusted codebase.

Programmers must audit not only any declassify operations or unsafe Rust code in their applications, but also the application's \lstinline{Cargo.toml} file, which specifies the names and locations of external crates including \ifclib and the Rust Standard Library.

\subsubsection*{Macros}
\label{par:macros}

What if application code calls an application macro that transforms a secret block (e.g., \lstinline|evil_macro!(secret_block!(...))|)? Fortunately, the Rust compiler's macro expansion algorithm iteratively expands macros, starting from the outermost macro.
The resulting expansion is repeatedly expanded, until no macros are left in the expansion.
If an application macro expands to contain a call to \ifclib's macro, \ifclib's macro's expansion is opaque to the application macro.

\Ifclib requires all accesses to secret values to occur in macros---except for calls to \lstinline{declassify()} methods, which could be transformmed by an application macro as described above. However, this security hole is mitigated by the fact that \lstinline{declassify()} calls must already be audited. Alternatively, one could modify \ifclib's programming model so that declassification uses a macro instead of a method.

\subsection{Limitations}
\label{subsec:limitations}

% \mike{TODO: Continue revising this part based on reviewer feedback, including making it clear that we're not claiming that something's possible unless we've done it.} \ada{Does this extend to verbiage like `we argue that it should be feasible to do X'? If not, this section will become a list of things we don't do with no argument as to how \ifclib can be extended to support them.}
%\mike{Revised to try not to assert anything we don't know to be true.}

\Ifclib's design
% demonstrates the potential for providing strong IFC guarantees for an off-the-shelf mainstream imperative language. However, the current design
has some drawbacks that limit its practicality as a fine-grained IFC approach.
It may be possible to remove these limitations in future work.
% We argue that these limitations---which generally fall into two categories, design limitations and implementation limitations---are not insurmountable and can be addressed with more engineering effort.

% \subsubsection*{Design limitations}

\Ifclib's design supports only static secrecy labels, not integrity labels or dynamic labels (i.e., labels as run-time values).
% Dynamic labels in particular are needed for application data for which the secrecy label is not known until run time.
% We are not aware of any properties of \ifclib that would preclude adding support for dynamic labels and integrity labels while continuing to meet the requirements described in \autoref{subsec:requirements}, except that dynamic labels would impact run-time performance.
% Adding integrity labels to \ifclib, which are the dual of secrecy labels, should be straightforward. Adding dynamic labels would be a significant change since it would add run-time tracking and checking of dynamic labels. With dynamic labels, \ifclib's rudimentary effect system would still provide the needed guarantees, by preventing unchecked explicit and implicit flows.

% \subsubsection*{Implementation limitations}
\label{subsec:impl-limitations}

The current design does not allow the use of overloaded operators in secret blocks.
% With more engineering effort, it should be possible to allow programs to mark overloaded operators as \lstinline{[side-effect-free]} and have them checked by a \ifclib macro.
Likewise, the current design disallows application types that implement custom dereference (\lstinline{Deref} and \lstinline{DerefMut}) and destructor (\lstinline{Drop}) operations. We believe that this is not very restrictive because these traits are most useful for implementing abstract data types (ADTs). In fact, the Rust documentation\footnote{\url{https://doc.rust-lang.org/1.23.0/std/ops/trait.Deref.html}} emphasizes that \lstinline{Deref} should only be implemented by programmers when defining smart pointers. As a result, only a few ``kernel'' ADTs are needed for most software.
% and supporting these kernel ADTs is sufficient in most cases.
\ignore{
With more engineering effort, it may be possible to allow programs to define and use custom overloaded operators and dereference and destructor operations in secret blocks.
}%

Secret blocks cannot use macros. Due to the Rust compiler's macro expansion algorithm, which expands the outermost macro first and iteratively works its way through the resulting expansion (\autoref{subsec:security-guarantees-threats}), \ifclib cannot inspect a macro's expansion within a secret block and thus cannot certify that it is side effect free. It may be possible to extend \ifclib's macro to recognize known macros such as \lstinline{vec!}
% and even \lstinline{secret_block!()} (to support nested secret blocks, which \ifclib does not currently support)
and generate code using the appropriate macro expansion, but we have not implemented this.

Secret blocks and side-effect-free functions cannot use interior mutability.
In our experience, interior mutability is common, but is not typically necessary:
In the evaluated programs, we were able to refactor the code in all side-effect-free code to not use interior mutability.
% \mike{Reviewer says: ``in my experience, interior mutability is quite common! But I would be willing to believe that it's not necessary for most secret blocks.'' Revised.}
% \ada{Removed reference to Spotify-TUI case study since it uses \lstinline|Result| which is supposedly interior mutable? TODO: confirm.}
% \mike{\lstinline|Result| is super common. Is \lstinline|Result| interior mutable externally (i.e., can applications mutate the interior of an immutable \lstinline|Result|) or just internally?
% Also, we add zero secret blocks to Servo, so the above statement about Servo is vacuously true. :)}
% \mike{This isn't actually problem. See my updated comments in Eval.}

Calls from side-effect-free code to allowlisted Rust Standard Library functions must be fully qualified (\autoref{subsec:prohibiting-calls-with-side-effects}). As an alternative that we have not implemented, \ifclib's procedural macros could rewrite every non-fully qualified call that \emph{syntactically appears to be} a call to a Rust Standard Library function, to a fully qualified call. For example, \lstinline{s.trim()} would be rewritten as \lstinline{::std::string::String::}\allowbreak\lstinline{trim()}.
% A future implementation could eliminate this requirement by automatically rewriting function calls to match their fully qualified names.
To avoid ambiguity of function names, \ifclib could disallow side-effect-free application functions from having the same name as any allowlisted library function.
    \section{Evaluation}
\label{sec:eval}

This section evaluates the usability and performance of \ifclib.
% \mike{Reviewer says: ``the evaluation states that the paper is evaluating `usability'. That word then never appears again. If I ask my self, `Section~\ref{sec:eval} shows that Cocoon is usable because: \_\_\_\_\_' then I'm not 100\% sure how I'm expected to fill in the blank. Because it's possible to use it within big Rust systems? Because the changes to the system are small? The paper should explicitly spell this out somewhere.'' Revised this paragraph.}
We integrated \ifclib with two applications (Spotify TUI and Servo) and also wrote \ifclib and non-\ifclib versions of a Battleship program.
These case studies show the usability of \ifclib: \Ifclib can be used by real applications to enforce security policies, with modifications commensurate with the amount of code that deals with secret values.
% finding that applications can start reaping the benefits of the library with limited modifications.
% ~\cite{Jif,JRIF} to Rust using \ifclib.
A performance evaluation shows no detectable impact on executable size or run-time performance,
and modest compile-time overheads for real applications.
% To evaluate performance, we evaluated the worst case for performance---all application code is marked side effect free---and found that
% In an evaluation of worse-case performance,
% We evalauted worse-case performance,
% To evaluate worst-case performance, we present performance results for \ifclib using a set of performance benchmarks (Rust programs from the Benchmarks Game~\cite{benchmarks-game}),
% finding that
% \ifclib increases compile times by 7--53\% but has no detectable impact on executable size or run-time performance.
% The evaluation finds no statistically significant performance differences between the original programs and their \ifclib-retrofitted counterparts. \Ifclib does however increase compile times.

% \paragraph{Platform}

% For all evaluated programs except Servo,
All experiments were executed on a quiet machine
with a 16-core Intel Xeon Gold 5218 at 2.3\,GHz with 187\,GB RAM running Linux.
% We ran Servo's experiments on a different machine---an 8-core Intel Xeon E5405 at 2 GHz with 16 GB RAM running Linux---because of difficulty we encountered in replicating Servo's build and run-time environment.

\subsection{Case Study: Spotify TUI}
\label{subsec:eval-spotify}

Spotify TUI (\underline{t}ext-based \underline{u}ser \underline{i}nterface) is a Spotify client written in Rust that allows an end user to interact with Spotify from a terminal. We chose it as a case study because it is a popular project on GitHub\footnote{\url{https://github.com/Rigellute/spotify-tui}}
%\mike{For URLs, put them in a footnote (or cite a reference if/when we're running into the page limit, since refs don't count toward page limit)}
with over 13,000 ``stars.''
%In fact, one of the authors used it as their music player for a few months.
%
%\mike{This paragraph jumps in without it being clear why we're telling the reader what we're telling them.}
%
%Spotify TUI is a complex program with over 12,000 lines of source code and over 90 contributors. For this study, we used the latest commit (\code{c4dcf6}) available at the time of writing.
%\mike{Which version/commit (including date) of Spotify TUI are we using?}

We modified Spotify TUI to use \ifclib to protect a value called the \emph{client secret}, which is akin to a user password. The client secret is a 32-digit hexadecimal string that Spotify TUI uses to authenticate with the Spotify API server.
% \autoref{fig:spotify-tui-client-secret} gives the original and \ifclib definitions of the \textit{client secret}.
%\mike{I disagree with including the definitions of the struct in this way. \Ifclib protects values, not memory locations like \lstinline{ClientConfg::client_secret}. Furthermore, the client secret flows between different memory locations in Spotify TUI's code---it doesn't just live in \lstinline{ClientConfg::client_secret}. I think the figure and the sentence referring to it should be removed. (Although the paper shows a similar figure for Servo, Servo is a special case: In Servo we're really just using \ifclib to protect a secret location, not a secret value.)
%Update: Removed the struct definitions.}%
Leaking the client secret to an adversary would allow the adversary to make Spotify API calls on behalf of the victim user.

\iffalse
\begin{figure}[t]
    \begin{minipage}[t]{.33\textwidth}
\begin{lstlisting}(*@\label{lst:client-secret-orig}@*)
pub struct ClientConfig {
 pub client_id: String,
 pub client_secret: String,
 pub device_id: Option<String>,
 pub port: Option<u16>,
}
\end{lstlisting}
    \centering \small \textsf{Original definition.}
    \end{minipage}
    \hfill
    \begin{minipage}[t]{.57\textwidth}
\begin{lstlisting}(*@\label{lst:client-secret-cocoon}@*)
pub struct ClientConfig {
 pub client_id: String,
 pub client_secret: sec::Secret<String, lat::Label_A>,
 pub device_id: Option<String>,
 pub port: Option<u16>
}
\end{lstlisting}
    \centering \small \textsf{Definition after porting to use \ifclib.}
    \end{minipage}
    \caption{Definitions of the \textit{client secret} in Spotify TUI before and after modification to use \ifclib. \ada{Should I include the derive statement for the structs? They differ between Original and \Ifclib since we can't derive Serialize, PartialEq, or Deserialize} \mike{I'm confused---what does the actual code look like?} \mike{See my comment in the text about what this figure shows/suggests.}}
    \label{fig:spotify-tui-client-secret}
\end{figure}
\fi

% In the case of Spotify TUI, each end user is responsible to acquire their own client ID and client secret, and, though Spotify TUI can control which track is being played, the actual streaming is done via a separate client, such as the official Spotify client.

%A user obtains the client secret by registering on the \emph{Spotify for Developers} web page.
% \footnote{\url{https://developer.spotify.com/}}.
Spotify TUI receives the client secret from the command line, or from a file that a previous execution of Spotify TUI wrote the client secret to. We modified these code locations to wrap the client secret as a \code{Secret<String,Label_A>}.
(For this application we use a simple label scheme in which $\{a\}$ and $\emptyset$ are the only labels, representing secret and non-secret values, respectively.)
Spotify TUI supports sending the client secret outside of the application in two ways: (1) writing the client secret to a file; and (2) sending the client secret to two external Rust libraries, RSpotify (a Rust wrapper for the Spotify Web API) and Serde (a Rust serialization library).
%\begin{itemize}
%\item It writes the client secret to a file as mentioned above.
%\item It sends the client secret to two separate Rust libraries: RSpotify, a Rust wrapper for the Spotify Web API, and Serde, a Rust serialization library.
%\end{itemize}

By retrofitting the Spotify TUI application to use \ifclib, we implicitly treat Spotify TUI as untrusted (except for its declassify calls and any \code{unsafe} blocks). Spotify TUI must declassify the client secret in order to write it to a file and to send it to RSpotify and Serde (because we have not integrated these components with \ifclib), implicitly treating the OS, RSpotify, and Serde as part of the trusted codebase. Removing RSpotify and Serde from the trusted codebase would require retrofitting them with \ifclib by annotating their side-effect-free functions (\autoref{subsec:security-guarantees-threats}).

\begin{figure}[t]
    \centering
    \lstset{numbers=left,basicstyle=\ttfamily\scriptsize}
    \begin{lstlisting}
fn validate_client_key(key: &str) -> Result<()> {
    if key.len() != EXPECTED_LEN {
        Err(Error::from(std::io::Error::new(
            std::io::ErrorKind::InvalidInput,
            format!("invalid length: {} (must be {})", key.len(), EXPECTED_LEN,),
        )))
    } else if !key.chars().all(|c| c.is_digit(16)) {
        Err(Error::from(std::io::Error::new(
            std::io::ErrorKind::InvalidInput,
            "invalid character found (must be hex digits)",
        )))
    } else {
        Ok(())
    }
}
\end{lstlisting}
\caption{Spotify TUI code that validates that the client secret is a 32-digit hexadecimal string.}
%    {\small \sf (a) Original code}
\label{fig:Spotify-Validation-Original}
    %\label{fig:Unmodified_Validation}
%\end{figure}
\bigskip\bigskip
%\begin{figure}[htbp]
    \centering
    \lstset{numbers=left,basicstyle=\ttfamily\scriptsize}
    \begin{lstlisting}
fn validate_client_key(key: &Secret<String, Label_A>) -> Result<()> {
  const EXPECTED_LEN: usize = 32;
  let sec_error_string = secret_block!(Label_A {
    let u_key = unwrap_secret_ref(key);
    let mut is_hex = true;
    for c in ::str::chars(&u_key) {
      if !::char::is_digit(c, 16) {
        is_hex = false;
      }
    }
    let mut error_string = ::std::string::String::from("");
    if ::std::string::String::len(&u_key) != EXPECTED_LEN {
      error_string = /* create string "invalid length: {key.len()} (must be {EXPECTED_LEN})" */;

    } else if !is_hex {
      error_string = ::std::string::String::from("invalid character found (must be hex digits)");
    }
    wrap_secret(error_string)
  });

  let error_string = sec_error_string.declassify();
  if !error_string.is_empty() {
    Err(::std::error::Error::from(::std::io::Error::new(
      ::std::io::ErrorKind::InvalidInput,
      error_string,
    )))
  } else {
    Ok(())
  }
}\end{lstlisting}
%    {\small \sf (b) \Ifclib-retrofitted code}
    %\\\mike{Code needs some changes to match the \ifclib usage in the rest of the paper.} \ada{I updated the labels and declassify call - that looks like the only necessary changes to me} \mike{Yeah I think that's it. Can you check the next two figures?} \ada{yep}
    \caption{Modified version of \autoref{fig:Spotify-Validation-Original}'s that uses \ifclib.}
    \label{fig:Spotify-Validation-Cocoon}
\end{figure}

% Commented out from if block in above figure
%::str::to_string("invalid length: ")
%        + &*::usize::to_string(&::std::string::String::len(&u_key))
%        + &*::str::to_string(" (must be ") + &*::usize::to_string(&EXPECTED_LEN)
%        + &*::str::to_string(")");

%\subsubsection*{Code excerpts}

As an example,
% Next we show a snippet of Spotify TUI code before and after retrofitting it to use \ifclib.
% to make the client secret be a \code{Secret<String,Label_A>}.
%\ada{TODO: Rewrite to remove \autoref{fig:Spotify-Validation} and \autoref{fig:Spotify-Write} I don't think these changes are particularly relevant. We should instead present the modification to \lstinline|client_secret|, not how it's used. Readers will have access to the code if they want to see the code in \autoref{fig:Spotify-Validation}. This change will bring the spotify-tui section in line with servo and reduce space.}
%
%\paragraph{Validation of the client secret}
%
\autoref{fig:Spotify-Validation-Original} shows a Spotify TUI function that validates that the client secret
% an input ``key,'' which could be either the client ID or client secret,
is 32 hexadecimal digits, returning a \code{Result} with value \code{Ok} or \code{Err}.
\autoref{fig:Spotify-Validation-Cocoon} shows how we changed the code to account for the client secret being a \code{Secret} value.
% The retrofitted code shows a differently named function, \code{validate_client_secret}, which it uses only for validating the client secret; it continues to use \code{validate_client_key} for the (non-secret) client ID, which has the same format requirements.
%\mike{I thought the differently named functions was a confusing detail, so I removed it and gave the function the same name in the figure. Note that we could actually make the presented code work by assigning a different name to the non-\ifclib version of the function.} \ada{sounds good.}
% retrofitted with \ifclib after marking the client secret as a achieve the same result for the client secret, a \code{Secret} type.

To perform the validation logic,
%the retrofitted code creates a secret block that processes the block's code with a procedural macro.
the modified code creates a secret block to check that each digit within the client secret is a hexadecimal character. Note that we had to manually implement the string formatting and hexadecimal character check, as opposed to using \code{|c| c.isdigit(16)}, since the \ifclib implementation does not support
% function or
% \mike{?}
closure calls within secret blocks.

%Because our prototype implementation's procedural macro does not currently support functions containing macro calls or closure invocations, we had to write the retrofitted code to perform the logic using expanded code that does not call the macro \code{format!} or use the inlined closure (\code{|c| c.isdigit(16)}). %The secret block calls only allowlisted side-effect-free functions in the Rust Standard Library; it always uses the fully qualified name (e.g., \code{::std::string::String::len}) to eliminate any ambiguity (\autoref{sec:application-defined-functions-visible}).

% In theory, the block could return a \code{Result}. However, the implementation of \code{Result} uses interior mutability, so \code{Secret<Result,Label_A>} is disallowed (\autoref{sec:application-defined-functions-visible}).
% Instead, we wrote
%\mike{Correction to text that was previously here: \lstinline{Result} isn't interior mutable. Or at least I was able to use \lstinline{Result} in a secret block just by implementing \lstinline{InvisibleSideEffectFree} for \lstinline{Result}. So I think using \lstinline{Result} in this case would be an option. But for now I revised the text to just say what the code does, not that we're avoiding using \lstinline{Result}.}%
The secret block populates an error string, which it leaves empty in the event of no error, and
%The retrofitted code passes the string as the mutable parameter to \code{apply_ternary_mut_ref}.
the retrofitted code declassifies the error string because, in the event of an error, it must report the error. Declassifying the error string is acceptable leakage because knowing \emph{whether} the client secret is correctly formatted is not security sensitive. The retrofitted code in \autoref{fig:Spotify-Validation-Cocoon} ultimately returns a \code{Result}, just as in \autoref{fig:Spotify-Validation-Original}. Thus, \autoref{fig:Spotify-Validation-Cocoon} achieves the same behavior for the client secret without declassifying the secret---but rather by declassifying only whether the input client secret is not in the correct format, which is not exploitable information.

\iffalse
\begin{figure}[t]
    \centering
    \lstset{numbers=left,basicstyle=\ttfamily\scriptsize}
    \begin{lstlisting}
let mut oauth = SpotifyOAuth::default()
    .client_id(&client_config.client_id)
    .client_secret(&client_config.client_secret)
    .redirect_uri(&client_config.get_redirect_uri())
    .cache_path(config_paths.token_cache_path)
    .scope(&SCOPES.join(" "))
    .build();
...
request_token(spotify_oauth);
    \end{lstlisting}
    {\small \sf (a) Original code}
    %\label{fig:Unmodified_OAuth}
%\end{figure}
\bigskip
%\begin{figure}[htbp]
    \centering
    \lstset{numbers=left,basicstyle=\ttfamily\scriptsize}
    \begin{lstlisting}
let mut oauth: Secret<SpotifyOAuth, Label_A>  = Secret::new(
    SpotifyOAuth::default()
        .client_id(&client_config.client_id)
(*@\label{fig:Spotify_TUI_Case_Study_OAuth}@*)        .client_secret(client_config.client_secret.declassify_ref())
        .redirect_uri(&client_config.get_redirect_uri())
        .cache_path(config_paths.token_cache_path)
        .scope(&SCOPES.join(" "))
        .build()
);
...
request_token(spotify_oauth.declassify_ref());  
    \end{lstlisting}
    {\small \sf (b) \Ifclib-retrofitted code}
    \caption{Spotify TUI code in which the client secret is passed to a \code{SpotifyOAuth}, which is later used to request a token.}
    \label{fig:Spotify-OAuth}
\end{figure}

\paragraph{Sending the Secret to RSpotify}

\autoref{fig:Spotify-OAuth}(a) shows unmodified code in which the client secret is passed to a \code{SpotifyOAuth} and then that \code{SpotifyOAuth} is used, in a different part of the code, to request a token.  \autoref{fig:Spotify-OAuth}(b) shows retrofitted code with the same behavior except that the \code{SpotifyOAuth} is made \code{Secret}. The code makes the call to \code{declassify\_ref} on line \ref{fig:Spotify_TUI_Case_Study_OAuth} because \code{SpotifyOAuth} is defined in the RSpotify codebase, which has not been retrofitted to use \ifclib, so not doing so will cause a compile error.  The \code{SpotifyOAuth} must be made \code{Secret} to prevent it from leaking because it contains the client secret.  \autoref{fig:Spotify-OAuth}(b) serves to demonstrate how the client secret is used to request a token in the retrofitted code.
\fi

\iffalse
\begin{table}[h]
    \centering
    \caption{The number of times the retrofitted code called one of \ifclib's \lstinline{Secret::declassify} methods, and reason for each call.}
    \begin{tabular}{rrr|r}
        \hline
        \textbf{Secret leaves codebase} & {\textbf{Acceptable leakage}} &
        {\textbf{Secret calls method}} & {\textbf{Total}} \\
         %& Original & \ifclib & Original \\
        \hline
        13 & 3  & 1 & 17  \\
        \hline
    \end{tabular}
    %\label{tab:spotify_declassify_methods}
\end{table}
\fi

\subsubsection*{Empirical Results}
\label{subsec:spotify-results}

\autoref{tab:code_changes} shows that the annotation burden of protecting the client secret with \ifclib is small: To retrofit Spotify TUI, we inserted 167 lines of code and removed 54 lines in a codebase of over 12,000 lines.

\autoref{tab:code_changes}'s last column shows how many times a call to a \lstinline{Secret::declassify} method appears in the retrofitted code of Spotify TUI. In 14 out of 17 cases, the code declassifies secret data because the data needs to be sent outside of the Spotify TUI codebase. In these cases, the client secret leaves the codebase to be passed to RSpotify or Serde, or to be written to the configuration file, all as intended. In the other 3 out of 17 cases, the code declassifies values that do not consitute a secrecy leak. For example, revealing the \lstinline{Result} in \autoref{fig:Spotify-Validation-Cocoon} does not present a security risk.

\iffalse
\autoref{tab:spotify_declassify_methods} shows the reasons that declassify methods were used in the retrofitted code and the number of times declassify was called for each reason.  Acceptable leakage refers to any situation in which a value other than the client secret is made secret, e.g., it needed to be mutated in a secret closure, but we decided the value can be declassified. In one case, a secret was the receiver of a method defined outside the codebase, so declassify needed to be called first to call that method.  The most important metric in evaluating the ability of the library to be used to audit information flow is the percentage of times declassify was called because the secret left the codebase to the total. With a large value of 76\% (13/17), it indicates that a developer can easily observe the uses of declassify and quickly gain an understanding of where the secret is leaving the codebase.  We believe this success is attributable to the expressiveness of our secret closures.  In this case, the client secret leaves the codebase to be passed to RSpotify, to be passed to a serialization library, and to be written to the configuration file, all as intended. If declassify methods were used often for reasons other than to allow the secret to leave the codebase, then calls to declassify would not provide good insight to the information flow.
\fi

\begin{table}[t]
\small
\centering
\caption{Programmer burden for integrating three applications with \ifclib, in terms of lines of code modified, and number of calls to declassify methods.}
\begin{tabular}{@{}l|rrrr|r@{}}
& \multicolumn{4}{c|}{Lines of source code} & \\
Program & Original & \Ifclib & Inserted & Deleted & Declassify calls \\\hline
 Spotify TUI & 12,169 & 12,282 & 167 & 54 & 17 \\
Servo & 397,141 & 397,174 & 77 & 44 & 1 \\
Battleship & 383 & 410 & 47 & 20 & 2
\end{tabular}
%\\\mike{(As suggested on Slack) can you verify visually/manually that the diff is really as reported, i.e., none of the insertions/deletions are due to trivial/unnecessary differences?}\ada{Updated spotify-tui to reflect diff without space. See thread on slack. The diff itself has been manually checked}
\label{tab:code_changes}
\bigskip\bigskip
\newcommand\mypm{\ensuremath{\pm}}
\setlength{\tabcolsep}{1.9pt}
\small
\caption{Comparison of performance costs for applications without \ifclib and versions that use \ifclib. Execution times include 95\% confidence intervals based on multiple trials.}
\begin{tabular}{@{}l|rr|rr|rr@{}}
%\begin{tabular}{@{}l@{\;}|@{\;}c@{\;\;}c@{\;}|@{\;}c@{\;\;}c@{\;}|@{\;}c@{\;\;}c@{}}
& \multicolumn{2}{c|}{Compile time (s)} & \multicolumn{2}{c|}{Run time} & \multicolumn{2}{c@{}}{Executable size (bytes)} \\
Program & Original & \Ifclib & Original & \Ifclib & Original & \Ifclib \\\hline
Spotify TUI & 14.395 \mypm{} 0.031 & 14.395 \mypm{} 0.053 &  2868.711\,\mypm{}\,5.612\,ns  & 2822.784\,\mypm{}\,5.662\,ns & 15,779,320 & 15,779,088 \\
% Spotify TUI & 15.977 \mypm{} 0.039 & 16.072 \mypm{} 0.059 &  101.665 \mypm{} 4.802$\,\mu$s  & 102.281 \mypm{} 2.722$\,\mu$s & 5,570,400 & 5,561,048 \\

% Old spotify results (11/02/2022)
%Spotify TUI & 145.9\mypm{}0.406 & 148.5\mypm{}0.968 &  316.717\mypm{}3.825$\,\mu$s  & 315.633\mypm{}3.908$\,\mu$s & 2,042,096 & 2,046,848 \\

%Servo & 1,268.764\mypm{}13.620  & 1,269.045\mypm{}13.677 & 22.549\mypm{}0.046\,s & 22.551\mypm{}0.032\,s & 370 & 370
% Servo & \textcolor{red}{1,268.764\mypm{}13.620}  & \textcolor{red}{1,269.045\mypm{}13.677} &\textcolor{red}{ 22.549\mypm{}0.046\,s} & \textcolor{red}{22.551\mypm{}0.032\,s} & \textcolor{red}{369,532} & \textcolor{red}{369,560}
Servo & 312.579\,\mypm{}\,4.354 & 309.983\,\mypm{}\,4.262 &183.817 \mypm{} 0.540\,s & 184.086 \mypm{} 0.529\,s & 403,548,416 & 403,547,984 \\
Battleship & 2.068 \mypm{} 0.003 & 2.321 \mypm{} 0.002 & 30.240 \mypm{} 0.254\,ms & 29.490 \mypm{} 0.272\,ms & 6,727,464 & 6,740,248 \\
%Spotify TUI & 145.9\mypm{}0.406 & 148.5\mypm{}0.968 &  316.717\mypm{}3.825$\,\mu$s  & 315.633\mypm{}3.908$\,\mu$s & 2,042,096 & 2,046,848 \\
%Servo & 1,268.764\mypm{}13.620  & 1,269.045\mypm{}13.677 & 22.549\mypm{}0.046\,s & 22.551\mypm{}0.032\,s & 369,532 & 369,560
\end{tabular}

%\mike{We aren't reporting memory usage for Spotify-TUI and Servo. One option is to remove memory usage from all tables and instead just say in the text that the performance benchmarks report memory usage and there's no statistically significant difference.}

%\emph{Exec.\ size} is the size of the machine code executable.}
\label{tab:perf}
\end{table}

\autoref{tab:perf} shows the performance cost of retrofitting Spotify TUI to use \ifclib. We compiled Spotify TUI 10 times, without and with integration with \ifclib, using the ``release'' optimization level. According to the results, \ifclib has no detectable impact on compile time. \Ifclib has a modest impact on the size of the compiled binary; in fact, \ifclib's code size is actually smaller than the original's by 232 bytes, for reasons that are unclear to us. In theory, the compiled code should be identical, but the additional code generated by \ifclib's macros could affect compiler decisions such as how aggressively to inline.

To measure run time, we timed just the part of the code---configuration and authentication---that deals with the client secret, in both the original and retrofitted Spotify TUI. As the results show, the measured time is around 3$\,\mu$s. We ran and timed this part of the code 100,000 times for each version of Spotify TUI to account for high run-to-run variability.
% In any case, we plan to address this measurement issue to be able to report run times with less variability. The reported results show that if there is any performance overhead added by the modified version of Spotify TUI, it is at most about 1\%.
The \ifclib version actually outperforms the original version by about 1\%. We are unsure of the reason for this (statistically significant and repeatable) performance difference. Like the executable code size differences, the performance difference could be the result of compiler decisions impacted by the extra code generated by \ifclib's macros.
% We confirmed that this result is repeatable, but are unsure if it is the result of the compiler actually generating
% The result shows no statistically significant run-time difference between the original and \ifclib, but it does \emph{not} show that the actual overhead is negligible ($<\,$1\%). However, the performance results for the rest of our evaluated programs have less variability, establishing that \ifclib's run-time overhead is negligible or nonexistent.

\subsection{Case Study: Servo}
\label{subsec:eval-servo}

Servo is Mozilla's browser engine written in Rust~\cite{servo}.
% A browser engine creates visual representations of resources on a web page.
We selected Servo as an evaluation subject for two reasons. First, Servo is the sixth-most popular Rust project on GitHub, with over 20,000 ``stars.'' Second, Servo is an influential project, e.g., Rust's early design was informed by Mozilla's experiences creating Servo. We modified Servo to use \ifclib to help enforce a security policy that prevents different JavaScript programs from reading each others' user data.

To make an HTTP request, JavaScript programs call the Fetch API, which the JavaScript engine implements by calling into Servo's implementation to perform the request and return a response. Critically, JavaScript programs are only allowed to read HTTP responses if the programs share the same origin web server, unless the response explicitly allows cross-origin sharing. Otherwise a malicious JavaScript program such as a malicious advertisement could read sensitive user data such as a password field from a login form from a trusted JavaScript program originating from a different web server. This policy is called the \emph{same-origin policy}. Responses from different origin servers are considered \emph{opaque} and cannot be read by JavaScript programs.

% \maxwell{This bug is not this compelling. It forgets to mark the response as opaque in the first place. It motivated enforcing the same-origin policy, but it cannot be patched using our library.}
% A prior Servo version had a security bug in its enforcement of the same-origin policy.\footnote{\url{???} \mike{TODO: link to bug report}} This bug has been fixed in the Servo version we modified (\lstinline{35e95f}), but we decided to try to fix the issue using \ifclib, which has the potential to be more robust and secure.

% JavaScript programs use the fetch library to make HTTP requests. A JavaScript engine (e.g., Mozilla's SpiderMonkey) uses Servo's fetch API to provide the fetch library to the JavaScript programs it executes. JavaScript programs are only allowed to read HTTP responses that share an origin server with the program unless the response explicitly allows cross-origin sharing. This prevents leaking user secrets (e.g., by a malicious advertisement reading the contents of a password field in a login form). This is called the \emph{same-origin policy}. Responses from different origin servers are considered \emph{opaque}. Opaque responses cannot be read by JavaScript programs.

%        unsafe {
%            (*@\label{fig:Servo_Case_Study_forget}@*)std::mem::transmute(Arc::clone(
%                &self.body))
%        }
\autoref{fig:Servo_Case_Study} shows how we modified Servo to use \ifclib. First, we modified the \lstinline{Response} type, which is instantiated after the client receives an HTTP response from an origin server following an HTTP request,
% We made the response body private to prevent other parts of Servo from reading the response.
% \mike{I think making the field private is a red herring here: Making the field private helped us identify the places where it was used, but now the fact that it's private isn't serving any purpose---being wrapped in a \code{Secret} protects it---and it can and should be changed back to public.} \maxwell{Marking it private means only code in this crate can access it. It lets us state that we \emph{know} if the response is marked opaque then the response body cannot be read.} \mike{Okay I see your point, but I disagree on making the field private: I think it doesn't make sense conceptually as part of retrofitting with \ifclib. Another point I'll make is that access modifiers protect variables, while IFC protects \emph{values}, and we should stay focused on the latter. Regardless of how we view this issue, saying we made the response body private to prevent other parts of Servo from reading the response seems potentially confusing to readers.}
to use \ifclib's \code{Secret} type to protect the response body from being read except by using declassification (lines~\ref{fig:Servo_Case_Study_before} and \ref{fig:Servo_Case_Study_after}).

\begin{figure}[t]
    \centering
    \lstset{numbers=left,basicstyle=\ttfamily\footnotesize}
    \begin{lstlisting}
pub struct Response {
(*@\label{fig:Servo_Case_Study_before}@*)(*@\diffdel{-\ \ pub body: Arc<Mutex<ResponseBody>{}>,}@*)
(*@\label{fig:Servo_Case_Study_after}@*)(*@\diffadd{+\ \ pub body: Arc<Mutex<Secret<ResponseBody,Label\_A>{}>,}@*)
  ...
}

(*@\label{fig:Servo_Case_Study_get_body}@*)pub fn get_body(&self) -> Arc<Mutex<ResponseBody>> {
    (*@\label{fig:Servo_Case_Study_opaque}@*)if self.response_type == ResponseType::Opaque {
        (*@\label{fig:Servo_Case_Study_empty_return}@*)Arc::new(Mutex::new(ResponseBody::Empty))
    } else {
        (*@\label{fig:Servo_Case_Study_forget}@*)self.body.declassify_transmute()
    (*@\label{fig:Servo_Case_Study_opaque_endif}@*)}
}
    \end{lstlisting}
    %\mike{Changed \lstinline|secrets::declassify_transmute(self.body)| to \lstinline|self.body.declassify_transmute()|. Please check.}
    \caption{Servo's response structure. We modified the structure to use \Ifclib to enforce the same-origin policy.}
    \label{fig:Servo_Case_Study}
\end{figure}
Following this change, the application can read the response body only by declassifying it.
% (or using an \code{unsafe} block).
To implement the same-origin policy, the code should only return the declassified value if the response type is not opaque, as shown in lines~\ref{fig:Servo_Case_Study_opaque}--\ref{fig:Servo_Case_Study_opaque_endif} of \autoref{fig:Servo_Case_Study}.
If the response is opaque, the code returns an empty body (line \ref{fig:Servo_Case_Study_empty_return}), in accordance with the Fetch API specification.
To return an \code{Arc<Mutex<ResponseBody>>}, the code calls a \code{declassify_transmute()} method (line~\ref{fig:Servo_Case_Study_forget}) provided by the \ifclib implementation; it is like other \lstinline{declassify} methods but returns a \code{T<Secret<U>>} value as a \code{T<U>}. \Ifclib implements \code{declassify_transmute()} safely without creating new data because \code{T<Secret<U>>} is guaranteed to have the same layout as \code{T<U>}.

Instead of adding this same logic at the dozens of locations in Servo's source code that read \code{Response::body}, we changed all such locations to instead call \code{Response::get_body()} (line~\ref{fig:Servo_Case_Study_get_body}).

Following this change, an opaque response's body is not readable without a declassify operation (or an \lstinline{unsafe} block). Our changes explicitly reflect developers' intentions and prevent future changes from accidentally allowing reads when the response body is opaque.
% \ada{The following sentence feels very subjective to me. What constitutes ``little'' burden? Considering the previous review that we need to be less subjective in our wording and allow readers to draw their own conclusions, we may want to delete this sentence.}
% \mike{Agreed}
% Overall, there was little burden in using \ifclib to enforce the same-origin policy.
%However, we did not use all of \ifclib's features in our enforcement; notably, \textcolor{red}{we did not use any secret blocks.}
%\mike{Probably isn't technically true anymore, now that secrets must be created with secret blocks, right?}

\subsubsection*{Empirical Results}

\autoref{tab:code_changes} (page~\pageref{tab:code_changes}) shows the effort to integrate Servo with \ifclib to protect the response body. We modified dozens of lines of code to integrate Servo with \ifclib. A significant number of the lines we modified reside in Servo's test suite (55). Many edits were made by our IDE's search-and-replace feature. At other source code locations, we had to introduce temporary variables because of Rust's lifetime analysis: Normally, the value returned by \lstinline{get_body} immediately dies (i.e., it is not usable by expressions later in the program). Storing the return value in a temporary variable keeps the value alive while the response body is read.

\autoref{tab:perf}
% (page~\pageref{tab:perf})
compares performance before and after integration with \ifclib. As for Spotify TUI, we compiled Servo 10 times, with and without integration with \ifclib. We observe that there is no statistically significant compile-time overhead imposed.\footnote{For Servo, compile time includes the time for compiling all dependent crates because Servo does not use the standard Rust build tools. For all of our other evaluated programs, compile time includes only the time to compile the target application.} We ran Servo's provided tests 20 times, with and without integration with \ifclib.
We find that \ifclib imposes no run-time burden that is statistically significant (the confidence intervals overlap). The executable code sizes differ by 432 bytes; as for Spotify TUI, the exact reasons are unclear to us because of the substantial amount of code involved and the complexity of the compiler.

\subsection{Case Study: Battleship}
\label{subsec:eval-battleship}

Battleship is a classic game in which two players place ships at secret locations on a grid. The object of the game is to sink the opponent's ships by correctly guessing their locations.
A player wins by sinking all of the other player's ships first.
% The first player with no ships remaining loses.

JRIF and Jif each employed Battleship as a case study~\cite{JRIF, Jif}. It is interesting as a case study not only because it requires confidentiality, but also because secret data (ship placements) are revealed throughout the game. An implementation of Battleship should not reveal any of the locations of a player's ships unless the opponent correctly guesses it first.  We wrote a multithreaded implementation of Battleship that uses \ifclib to enforce this secrecy policy.
%\maxt{Is it fair to say that we do not provide integrity?} \ada{Yes, we do mention above that Cocoon doesn't support integrity labels and its scope (for now) is on confidentiality only. } \maxt{From JRIF: ``For integrity, restrictions specify whether values may be considered trusted based on which principals might have influenced them. Principals should be trusted in order for the values they modify to be trusted.'' I do think that our implementation of Battleship provides this level of integrity. Meaning that we can trust that a player's ship locations are not modified by the other player in our implementation.}

\begin{figure}
  \centering
  \lstset{numbers=left,basicstyle=\ttfamily\scriptsize}
  \begin{lstlisting}
(*@\label{fig:game_state_ds_derive}@*)#[derive(InvisibleSideEffectFree)]
(*@\label{fig:game_state_ds_generic}@*)struct Player<L: Label> {
(*@\label{fig:game_state_ds_sp}@*)    ship_positions: Secret<Grid<bool>, L>,
    guesses: Grid<CellStatus>,
}
\end{lstlisting}
  \caption{Data structure representing each Battleship player's game state.}
  \label{fig:game_state_ds}
%\end{figure}
\bigskip\bigskip
%\begin{figure}
    \centering
    \lstset{numbers=left,basicstyle=\ttfamily\scriptsize}
\begin{lstlisting}
impl<L: Label> Player<L> {
    fn new() -> Player<L> {

        let ship_positions = (*@\label{fig:bs_init_ships}@*)secret_block!(L {
            let ships = [Ship::Carrier, Ship::Battleship, Ship::Cruiser, Ship::Submarine, Ship::Destroyer];
            let mut ship_positions: Grid<bool> = [[false; GRID_SIZE]; GRID_SIZE];
            for ship in ships {
                let placement: Placement = random_placement(&ship_positions, &ship);
                place_ship(&mut ship_positions, &ship, &placement);
            }
            wrap_secret(ship_positions)
        });

        Player {
            ship_positions,
            guesses: [[CellStatus::Unguessed; GRID_SIZE]; GRID_SIZE],
        }
    }
}
#[side_effect_free_attr]
fn random_placement(grid: &Grid<bool>, ship: &Ship) -> Placement {
    let mut ship_placement: Placement = random_maybe_illegal_placement(grid, ship);
    while !legal_placement(&grid, ship_placement) {
        ship_placement = random_maybe_illegal_placement(grid, ship);
    }
    ship_placement
}
\end{lstlisting}
    \caption{Battleship player initialization routines.}
    \label{fig:bs_game_init}
%\end{figure}
\bigskip\bigskip
%\begin{figure}
    \centering
    \lstset{numbers=left,basicstyle=\ttfamily\scriptsize}
\begin{lstlisting}
fn game_loop_a(mut player: Player<lat::Label_A>, chan: session_types::Chan<(), PlayerA>) {
    let mut c = chan.enter();
    let mut player_b_guesses = [[CellStatus::Unguessed; GRID_SIZE]; GRID_SIZE];
    loop {
        // Not shown: Display user's previous guesses, prompt user to make new guess, send guess to opponent.

        // Receive opponent's guess.
        (*@\label{fig:bs_loop_b_guess}@*)let (c3, guess) = c.recv();

        // Decide if the opponent hit our ship.
        (*@\label{fig:bs_loop_sb}@*)let is_hit: bool = secret_block!(lat::Label_A {
            wrap_secret(is_occupied(unwrap_secret_ref(&player.ship_positions), guess.0, guess.1))
        }).declassify();

        // Not shown: Update opponent's guesses. Exit loop if guess won the game.
    }
}

#[side_effect_free_attr]
fn is_occupied(grid: &Grid<bool>, row: usize, col: usize) -> bool { (&grid[row])[col] }
\end{lstlisting}
    \caption{Main Battleship game loop.}
    \label{fig:bs_game_loop}
\end{figure}

\autoref{fig:game_state_ds} shows the data structure that represents each player's game state. Line~\ref{fig:game_state_ds_derive} uses the \lstinline{#[derive(InvisibleSideEffectFree)]} macro to assert that the data structure is safe to use from within side-effect-free contexts. The struct uses a generic secrecy label (line~\ref{fig:game_state_ds_generic}) so that the \lstinline{Player} data structure can represent both players using distinct labels. Line~\ref{fig:game_state_ds_sp} shows how the \lstinline|ship_positions| field, which stores secret ship placement values, has \lstinline{Secret} type. The player's guesses are low-secrecy values since they are announced to their opponent, and thus the value of \lstinline{guesses} is not wrapped in \lstinline{Secret}.
% \mike{Reviewer says: ``try to wrestle with this paragraph and its surroundings so that its final sentence is not split across three pages''}
% \mike{Similar to the explanations of the other case studies, the discussion here is veering into suggesting that IFC protects the \lstinline{ship_positions} field, which is a memory location. That's not correct---rather, IFC protects the \emph{values} stored in that field. Those values can flow around to different expressions and variables. To be clear, the effect of making \lstinline{ship_positions} have \lstinline{Secret} type is \emph{not} to make the field or value secret or to protect the field or value, but rather to denote that the field can hold secret values. Update: I revised it a bit.}%

\autoref{fig:bs_game_init} shows the \lstinline{Player} initialization routine, which preserves the confidentiality of \lstinline{Player::ship_}\allowbreak\lstinline{positions}. Line~\ref{fig:bs_init_ships} begins the secret block that initializes a player's ship placements. The secret block iterates through the ships in the game and places them onto a grid.
% The \lstinline{#[side_effect_free_attr]} annotation on \lstinline{random_placement} checks the definition of \lstinline{random_placement} and rewrites it so that it can be called from other side-effect-free contexts.
The secret block calls \lstinline{random_placement}, which is annotated with \lstinline{#[side_effect_free_attr]}, demonstrating \ifclib's effect system.
% This allows us to call \lstinline{random_placement} without causing a compilation error (as would be the case in the absence of the annotation).
Note that \lstinline{random_placement} calls more application functions (\lstinline{legal_placement} and \lstinline{random_maybe_illegal_placement}) whose definitions are not shown.

The main game logic is shown in \autoref{fig:bs_game_loop}. The players communicate over the channel \lstinline{chan}. This code prints Player A's guesses (public information), then prompts Player A to make a new guess (public information). The code then sends the guess to Player B over \lstinline{chan}. Player B uses the same channel to communicate whether the guess hit a ship or ended the game. Next, Player A receives Player~B's guess (line \ref{fig:bs_loop_b_guess}). The game loop uses a secret block on line \ref{fig:bs_loop_sb} to decide if Player B scored a hit. This is necessary because the player's ship placements are secret.

\ignore{\ifclib enforces the same confidentiality policy as Jif \cite{Jif}. However, Jif supports integrity labels. Integrity labels are useful to prevent players from altering their ship placements after the game begins.}%

\subsubsection*{Empirical Results}

\autoref{tab:code_changes} compares two versions of our Battleship implementation: one with \ifclib, and one without it. The \ifclib changes affect a significantly larger fraction of the code than for Spotify TUI and Servo. There are two calls to \lstinline{declassify} in Battleship. Both occur in a player's game loop to decide if their opponent correctly guessed a ship location.

\autoref{tab:perf} shows the effect of \ifclib on compilation times, run times, and executable sizes, using the same methodology as for Spotify TUI and Servo.
% We compiled Battleship 10 times, without and with integration with \ifclib, using the ``release'' optimization level.
\Ifclib slows compilation time by 12\%---more than for Spotify TUI and Servo, which makes sense because a larger fraction of Battleship's code is in secret blocks and side-effect-free functions.
\Ifclib adds no run-time overhead---in fact, as for Spotify TUI, it decreases run time inexplicably, perhaps as a result of different compiler optimization decisions---showing that the compiler effectively optimizes away the type-checking scaffolding added by \ifclib. Finally, \ifclib increases executable size by 0.1\%, for unknown reasons, possibly again due to compiler optimization decisions.

% \mike{I removed the stuff about integrity and comparison with Jif.}
% \mike{It doesn't seem well justified that we don't have a non-\ifclib (Rust) version of Battleship.}
% \ada{TODO: Add comparison to non-Cocoon version of Battleship and add to \autoref{tab:code_changes} for battleship - no performance.}
% \mike{TODO: Make sure we've updated everything to include the addition of Battleship results.}

%\mike{\autoref{fig:bs_game_loop} seems like a lot to show, especially since almost none of it is in a secret block. Seems important to show \lstinline{is_occupied}; in any case it's very short. Maybe modify the code so it summarizes very briefly what happens before and after the secret block rather than showing it in depth?}

\subsection{Performance-Focused Evaluation}
\label{subsec:perf-eval}

To measure the worst-case performance impact of \ifclib, we evaluated wrapping entire programs in secret blocks. We used benchmarks from The Computer Language Benchmarks Game~\cite{benchmarks-game}, for which developers create the most efficient implementations possible, in order to compare the performance of different programming languages.
We evaluated 9 out of 10 available benchmarks~\cite{benchmarks-game}, excluding \bench{reverse-complement} because it concurrently modifies a shared mutable buffer, which constitutes an inherent side effect.

To wrap entire programs in secret blocks,
we placed the functions that perform the benchmark's primary calculations inside of a secret block, and annotated the blocks' transitive callees as side-effect-free functions.
We refactored all functions' code to adhere to \ifclib's programming model constraints.
We directed \ifclib's procedural macros to ignore calls to functions in external crates, by wrapping them in \lstinline{unsafe} blocks, which the procedural macro does not transform.

% We modified the best-performing Rust implementations to perform all of their computations with the primitives \ifclib introduces.
% In each case, we can conclude with high confidence that \ifclib adds negligible (<1\%) or nonexistent runtime overhead and increases compile time by 7--16\% with the exception of \lstinline|n-body|, which increased compile time by 53\%.
% For details of this evaluation, see Appendix~\ref{appendix:PLGame}.

\autoref{tab:PLGame} shows results of each benchmark with and without our modifications to use \ifclib. We observe that there are few statistically significant differences in run time (wall clock time)
%or memory usage (maximum resident set size)
between the original and \ifclib versions, and confidence intervals are tight enough that we can conclude with high confidence that \ifclib adds negligible ($<\,$1\%) or nonexistent overhead---and it speeds up programs as often as it slows them down, suggesting that statistically significant run-time differences are due to indirect effects such as compiler inlining decisions or microarchitectural sensitivities.
(We also measured run-time memory usage, i.e., maximum resident set size, but omitted the results from the table for brevity. The results show that space overhead is negligible or nonexistent for all programs, i.e., confidence intervals are consistently small and overlapping.)

\begin{table}
%\scriptsize
\small
\centering
\caption{Performance of nine benchmarks whose entire computation is wrapped in a secret block, compared with the originals. Performance results include the mean and a 95\% confidence interval from at least 10 trials.}
\begin{tabular}{@{}l|rr|cc|rr@{}}
%\begin{tabular}{@{}l@{\myspace}|@{\myspace}r@{\numspace}r@{\myspace}|@{\myspace}c@{\numspace}c@{\myspace}|@{\myspace}c@{\numspace}c@{\myspace}}%|@{\myspace}r@{\numspace}r@{}}
   & \multicolumn{2}{c|}{Run time (s)}
   %& \multicolumn{2}{c|@{\myspace}}{Max memory usage (KB)}
   & \multicolumn{2}{c|}{Compile time (s)}
   & \multicolumn{2}{c@{}}{Executable size (bytes)}\\
  Benchmark & \multicolumn{1}{c}{Original} & \multicolumn{1}{c|}{\ifclib} & \multicolumn{1}{c}{Original} & \multicolumn{1}{c|}{\ifclib} & \multicolumn{1}{c}{Original} & \multicolumn{1}{c@{}}{\ifclib} \\ %& Original & \Ifclib \\%\multicolumn{1}{c}{Original} & \multicolumn{1}{c@{}}{\ifclib} \\
  \hline
  \bench{binary-trees} & $0.364 \pm 0.023$ & $0.365 \pm 0.023$ %& $859,119 \pm 38,622$ & $804,017 \pm 45,910$
  & $0.608 \pm 0.010$ & $0.647 \pm 0.003$  & 4,560,608 & 4,560,640 \\
  \bench{fannkuch-redux} & $1.828 \pm 0.027$ & $1.821 \pm 0.011$ %& $3,412 \pm 40$ & $3,463 \pm 48$
  & $0.489 \pm 0.147$ & $0.545 \pm 0.002$  & 4,488,712 & 4,489,800  \\
  \bench{fasta} & $1.116 \pm 0.070$ & $1.093 \pm 0.049$ %& $2,676 \pm 57 & $2,699 \pm 60$
  & $0.417 \pm 0.008$ & $0.468 \pm 0.004$  & 4,339,184 & 4,339,640  \\
  \bench{k-nucleotide} & $1.539 \pm 0.007$ & $1.559 \pm 0.007$ %& $135,730 \pm 189$ & $135,904 \pm 272$
  & $0.464 \pm 0.003$ & $0.503 \pm 0.003$  & 4,355,808 & 4,360,544 \\
  \bench{mandelbrot} & $0.352 \pm 0.003$ & $0.352 \pm 0.003$ %& $31,597 \pm 256$ & $31,623 \pm 197$
  & $0.603 \pm 0.030$ & $0.699 \pm 0.006$  & 4,546,848 & 4,548,456 \\
  \bench{n-body} & $46.657 \pm 0.008$ & $46.335 \pm 0.005$ %& $2,067 \pm 44$ & $2,113 \pm 47$
  & $0.386 \pm 0.007$ & $0.591 \pm 0.007$ & 4,305,872 & 4,318,312 \\
  \bench{pidigits} & $0.554 \pm 0.005$ & $ 0.556 \pm 0.004$ %& $2,996 \pm 55$ & $3,036 \pm 44$
  & $0.334 \pm 0.004$ & $0.382 \pm 0.002$ & 4,528,080 & 4,529,800 \\
  \bench{regex-redux} & $0.957 \pm 0.003$ & $0.967 \pm 0.003$ %& $153,825 \pm 41$ & $153,811 \pm 42$
  & $ 0.644 \pm 0.005$ & $0.686 \pm 0.010$ & 6,114,664 & 6,115,904 \\
  \bench{spectral-norm} & $0.111 \pm 0.002$ & $0.112 \pm 0.003$ %& $3,922 \pm 52$ & $3,918 \pm 33$
  & $0.426 \pm 0.005$ & $0.477 \pm 0.005$ & 4,514,568 & 4,515,176 \\
\end{tabular}
\label{tab:PLGame}
\end{table}

\Ifclib increases compile time because its procedural macro adds significant ``scaffolding'' code for type checking, which is generally optimized away, so it does not translate to an increase in run time. The table shows that compile-time overhead ranges from 7\% to 16\% for all programs except \bench{n-body}, which has 53\% compile-time overhead. These overheads represent an extreme case in which all code is transformed by \ifclib's procedural macros. In practice, only a program's computations on secret values need to be transformed, which is why the compile-time overheads for Spotify TUI and Servo are significantly lower (\autoref{tab:perf}).

%\sout{To better understand potential compiled code differences, we examined the generated code for a simple example: \autoref{fig:Example_With_Secrets} and equivalent code without \ifclib's modifications. We found that the compiler generates identical code for both versions.}
%\mike{Removed this example from the paper. We could repeat with the new example (\autoref{fig:secure-cal}), but I think we can just omit this paragraph instead.} \ada{Sounds good.}

% \mike{How many trials of each experiment were run (or is it dynamic until the confidence interval is sufficiently small)?} \ada{The evaluate.sh and evaluate\_build.sh scripts in the benchmark games folder defines 10 trials. I updated the caption of table 3 to include this and be more consistent with table 2's caption.}

\iffalse
\paragraph{Compiled code differences}
%\ada{TODO: Update to be more standalone or move supplementary material back in.}
% \subsection{Impact on Compiled Code}
\label{subsec:compiled-code-impact}

% \begin{figure}
%     \begin{minipage}[t]{.45\textwidth}
%         \begin{lstlisting}
% (*@\label{fig:Simple_Code_Generation_1}@*)let mut secret_int =
%   sec::Secret::<i32, lat::A>::new(42);
% (*@\label{fig:Simple_Code_Generation_apply_unary_ref}@*)secret_int = sec::apply_unary_ref(
%   secret_macros::side_effect_free_closure!(
%     |x| x * 2
%   ),
%   &secret_int,
% );\end{lstlisting}
%     \end{minipage}
%     \hspace{.1\textwidth}
%     \begin{minipage}[t]{.45\textwidth}
%         \begin{lstlisting}
% mov     dword ptr [rsp + 12], 84\end{lstlisting}
%     \end{minipage}
%     \mike{I think this example is too contrived because 42 is hard-coded. Instead make a function that takes \lstinline{secret_int} as an input?}
%     \caption{The left listing shows a simple example of code written with \Ifclib. The right listing shows the code generated by the Rust compiler with a low optimization level.}
%     \label{fig:Simple_Code_Generation}
% \end{figure}

%\mike{Revised this section to be self-contained, so the reviewers don't need to read the appendix to follow it. (The appendix will be in a separate document that reviewers have to download separately. Most reviewers probably won't download it.)}

While our empirical results show that \ifclib adds no discernible run-time overhead and has little impact on compiled code size,
the evaluated programs are too large to understand the compiled code differences in detail.
% %
To better understand compiled code differences, we examined the generated code for a simple example: \autoref{fig:Example_With_Secrets} and equivalent code without \ifclib's modifications.
%The submitted supplementary material shows the x86-64 code.
\autoref{fig:Example_Without_IFC} shows the code in \autoref{fig:Example_With_Secrets} (page~\pageref{fig:Example_With_Secrets}) rewritten to \emph{not} use \ifclib. We compiled both versions of the functions with the Rust compiler's ``release'' optimization level. The compiler generates identical code for both versions, shown in \autoref{fig:Code_Gen}. Thus, common compiler optimizations are able to completely eliminate all traces of \ifclib from the application's final executable.

This result makes sense. The only member of the \lstinline{Secret} wrapper type is the wrapped value itself, so the size of \lstinline{Secret<T, L>} is always equal to the size of \lstinline{T}. Meanwhile, the Rust compiler performs function inlining, constant folding, dead code elimination, and other optimizations to eliminate the extra code added by \ifclib's procedural macro for compile-time checking.

\begin{figure}[t]
        \begin{lstlisting}
fn example(x: i64, y: i64) {
    let result = if x - y > 0 {
        "x wins!"
    } else {
        "x loses!"
    };
    println!("Result: {:}", result);
}
\end{lstlisting}
    \caption{Rust code that is equivalent to the code in Figure \ref{fig:Example_With_Secrets} (page~\pageref{fig:Example_With_Secrets}), but without using \ifclib.}
    \label{fig:Example_Without_IFC}
% \end{figure}
\bigskip
% \begin{figure}
    \centering
\begin{lstlisting}[mathescape=no]
example:
        subq    $88, %rsp
        subq    %rsi, %rdi
        xorl    %eax, %eax
        testq   %rdi, %rdi
        leaq    .L__unnamed_1(%rip), %rcx
        leaq    .L__unnamed_2(%rip), %rdx
        cmovgq  %rcx, %rdx
        setg    %al
        movl    $8, %ecx
        subq    %rax, %rcx
        movq    %rdx, 24(%rsp)
        movq    %rcx, 32(%rsp)
        leaq    24(%rsp), %rax
        movq    %rax, 8(%rsp)
        leaq    <&T as core::fmt::Display>::fmt(%rip), %rax
        movq    %rax, 16(%rsp)
        leaq    .L__unnamed_3(%rip), %rax
        movq    %rax, 40(%rsp)
        movq    $2, 48(%rsp)
        movq    $0, 56(%rsp)
        leaq    8(%rsp), %rax
        movq    %rax, 72(%rsp)
        movq    $1, 80(%rsp)
        leaq    40(%rsp), %rdi
        callq   *std::io::stdio::_print@GOTPCREL(%rip)
        addq    $88, %rsp
        retq
\end{lstlisting}
    \caption{x86-64 code generated by the Rust compiler for both \autoref{fig:Example_With_Secrets} and \autoref{fig:Example_Without_IFC}.}
    \label{fig:Code_Gen}
\end{figure}
\fi

    % related work
    \section{Related Work}
\label{subsec:related}

% pre-1970s - access control
% \maxt{I'm not a fan of this paragraph -- when do we use the information we share here? It's more like ``related work''. I feel it appears too early.} \ada{It's intended to give a history of IFC.}
% \mike{Moved to related work.}
%Historically, the focus of confidentiality research has been on access control,
% As defined by the U.S.\ Department of Defense's ``Orange Book'' \citeyearpar{DoD_85}, access control
%which restricts direct data access by annotating data labels with a secrecy label that defines who has access to this data~\cite{DoD_85}. Access restricts the inappropriate use of data but does not attempt to control indirect data access or the propagation of data after a valid initial access.
% 1970s
% Lampson - define confidentiality as not allowing the leak of even partial confidential data (modern IFC definition)
% Dennings with IFC and lattice, Dennings & Dennings with static analysis for IFC
In the mid-1970s, Lampson first discussed information confinement as preventing even the partial leak of confidential information~\citeyearpar{L_73}, while Rotenberg introduced a mechanism to constrain systems from allowing secret data to flow to untrusted entities~\citeyearpar{keeping_secrets}; these works became the concept of information flow control (IFC). Denning first defined a lattice structure for applying information flow policies~\citeyearpar{denning_lattices}, and Denning and Denning and later observed that static analysis can be used for IFC~\citeyearpar{DD_77}. Building on the static analysis models of the 1970s, Myers and Liskov introduced a decentralized model in which applications declassify their own data as opposed to relying on a third-party authority to do so~\citeyearpar{ML_97}.

% \mike{Mention that OS-level IFC is fundamental because a program's secrets come from outside the program and we're concerned when they flow outside the program (related to Max and Ada's questions in Intro).}
% \mike{Thinking about it more, language-level IFC without OS-level IFC isn't a toy---it's still useful. I plan to revise the various relevant places in the text to minimize language-level IFC's dependence on OS-level IFC.}
Prior systems have provided \emph{coarse-grained} IFC between processes, sockets, files, and other OS-level entities using dynamic tracking~\cite{dstar,flume}. In contrast, \emph{fine-grained} IFC at the level of variables and expressions presents unique challenges. First, communication within a process, which involves the flow of data and control between program-level elements such as values and expressions, is less explicit than the flow between processes and other OS-level entities, making it difficult to track information flow soundly and precisely. Second, fine-grained flow has higher bandwidth, so the cost of tracking it at run time can be substantial.

Existing fine-grained IFC operates either dynamically at run time or statically at compile time.

%Thus \emph{dynamic analysis} adds significant run-time overhead (as well as memory overhead to store per-value labels), and it is not well suited to handling implicit flows. On the other hand, \emph{static analysis} adds no memory overhead, but sound whole-program static analysis is imprecise, leading to many false violation reports. In contrast, using \emph{static typing}, programs specify types for variables and expressions that represent their secrecy labels, enabling precise compile-time checking of each function in isolation. However, existing typing approaches do not support mainstream imperative languages: They either support a functional language, or they extend an existing imperative language and thus rely on new syntax and development tools.

% Analysis-based IFC approaches attempt to check an entire program for IFC violations at run time or by using a standalone whole-program static checker.

\subsection{Dynamic IFC}

Dynamic IFC tracks and checks labels of variables and values at run time.
%Dynamic analysis can be precise (no false leak reports), but it adds run-time overhead. Another drawback is that dynamic analysis is not well suited to handling implicit flows (since an execution does not reflect all of the dependencies in the code). Furthermore, dynamic analysis means checks may fail unexpectedly at run time, hurting availability and potentially leaking information via termination channels.
%
Austin and Flanagan showed how to speed up dynamic tracking of labels for JavaScript programs, but their approach still adds significant run-time overhead~\citeyearpar{AF_2009}. Laminar tracks labels at run time for Java programs and handles implicit flows by consigning secret operations to lexically scoped regions, but still adds up to 56\% run-time overhead~\cite{laminar}. Co-Inflow provides dynamic coarse-grained IFC for Java programs through the use of implicitly inserted labels~\cite{coInflow}. Converting a Java program to use Co-Inflow requires few annotations and achieves relatively high precision, but introduces an average run-time overhead of 15\%.
% and does not support all Java features, such as reflection.
% \mike{I'm not sure we should sell this point.}

% In 2009, Austin and Flanagan presented a new, purely dynamic approach for IFC in JavaScript \citeyearpar{AF_2009}. They use sparse labeling to assign explicit labels for values that change security domains and use implicit labels for all others. This approach provides a 30--50\% speedup over universal labeling (in which all labels are explicit) but still costs around a second of run-time overhead per 100K program runs. In the same year, Roy et al.\ presented Laminar, the first decentralized information flow control system for Java using only a single set of abstractions to represent both OS resources and heap objects \citeyearpar{laminar}. Laminar uses a dynamic approach with lexically scoped regions to handle implicit flows and reduce run-time performance impacts but still incurs 1--56\% performance overheads.
%
% In 2021, Xiang and Chong introduced Co-Inflow, which extends Java to provide coarse-grained dynamic information flow control through the use of implicitly inserted labels \citeyearpar{coInflow}. Converting a Java program to use Co-Inflow requires few annotations while achieving relatively high precision but introduces an average run-time overhead of 15\% and does not support all Java features, such as reflection.

\subsection{Static IFC}

Static IFC can be provided by whole-program analysis or modular type-based analysis.

\subsubsection*{Whole-Program Static Analysis}

Whole-program static analysis computes information flow conservatively by abstracting data and control flow across program expressions and variables~\cite{JavaPDGs,SAILS,BBBPRR_17,HS_12}. It minimizes programmer effort by eschewing the need for type annotations, but it is non-compositional: Precision loss analyzing one part of the program affects precision analyzing other parts of the program. Non-compositionality makes it hard to scale to large programs and hard to deploy incrementally.
% minimizing run-time overhead and programmer effort, but reporting false flows that are difficult to deal with~\cite{JavaPDGs,SAILS,BBBPRR_17,flowistry,HS_12} (\autoref{subsec:intro-static-analysis}).
%Static analysis thus adds no run-time overhead and can handle implicit flows and avoid termination channels. However, sound whole-program static analysis is inherently conservative because it abstracts control and data flow, leading to false leak reports---which can be as time consuming to investigate as true leak reports, thwarting adoption. Prior work has extended advanced static analyzers~\cite{SAILS, BBBPRR_17} and introduced new path condition--based techniques~\cite{JavaPDGs} but still reports false leaks except on small test programs.

Recent work introduces whole-program static analysis for information flow in Rust called Flowistry~\cite{flowistry}. %Flowistry leverages the information inherent in function signatures and Rust's ownership guarantees to perform modular analysis (i.e., one function at a time) that is relatively precise. In essence,
Flowistry leverages Rust's ownership constraints to compute relatively precise function summaries without actually analyzing the functions' bodies. As such, Flowistry is scalable enough to use interactively, and it naturally handles calls to functions for which source code is unavailable. However, Flowistry is inherently \emph{intraprocedural}; Crichton et al.\ suggest that future work ``could build
an interprocedural analysis by using Flowistry's output as
procedure summaries in a larger information flow graph''~\cite{flowistry}.
As an intraprocedural analysis, Flowistry is not applicable to checking for illegal flows in our evaluated programs, which have secret values that flow interprocedurally.

As a static analysis approach, Flowistry may report false flows, which are hard for programmers to deal with.
In contrast, \ifclib uses static type-based analysis, which requires program modifications but provides a way to eliminate false illegal flows.
% In contrast, Cocoon uses static typing, so no separate tool is needed, and well-typed programs are guaranteed to compile (and have no illegal flows). Flowistry's contribution over prior static analysis--based approaches is that it leverages Rust ownership and lifetime properties to achieve impressive precision.

Flowistry and \ifclib are potentially complementary: Because Flowistry targets Rust and is fairly precise, it could potentially be adapted to infer some labels in applications using \ifclib.

\subsubsection*{Modular Static Type-Based Analysis}

Static type-based approaches extend the type system with IFC labels, so the static type of every expression and variable includes an IFC label.
In contrast with whole-program static analysis, static type-based analysis is compositional because typing rules can be checked locally (i.e., one expression at a time) instead of globally. As a result, type-based analysis offers the potential to allow large programs to provide IFC without false leak reports or run-time overhead.
A downside of type-based analysis is that it creates extra work for programmers to express types, although this burden can be limited to parts of the program that touch data with non-empty labels.
A key drawback of type-based analysis is that mainstream imperative languages such as Java, C, C++, C\#, and Rust do not provide type systems that are expressive enough to ensure that operations on secret values do not have unchecked side effects. As a result, existing solutions either use a functional language~\cite{MAC, DepSec} or extend an imperative language~\cite{JFlow, Jif, ML_97, JRIF, chong_dissertation, FlowCAML, spark, IFC_multithreaded, SV_type_model},
% by adding more advanced typing
as described in the following paragraphs.

MAC is a static type-based IFC library for Haskell that leverages the fact that operations in pure functional languages are side effect free~\cite{MAC}. Likewise, DepSec is a dependently typed static IFC library for Idris~\cite{DepSec}, a dependently typed language influenced by Haskell~\cite{Idris}. High-performance programs are typically not written in these languages.

% In 2019, Gregersen, Thomsen, and Askarov introduced DepSec, which provides a dependently typed library in Idris for static information flow \citeyearpar{DepSec}. DepSec is inspired by MAC \cite{MAC}, a statically enforced IFC library for Haskell, and leverages the fact that a pure functional language, like Haskell, is side effect free. Idris is a relatively young pure functional programming language developed by Brady in 2007 \citeyearpar{Idris}. It is strongly influenced by Haskell and has dependent types. Thus, DepSec aims to expand MAC to use full-spectrum dependent types, in which there is no limitation on the values that can appear in types. DepSec provides improved expressiveness over MAC but is less generally applicable to industry as Idris is less popular than Rust from a developer standpoint \cite{stackOverflow_survey}. %\ada{Re: DepSec. Not clear to me from reading the paper what DepSec doesn't solve that we do. As far as I can tell, the only advantage of \ifclib over DepSec is if the user already has a Rust system and doesn't want to change to Idris. I also don't know the nuances of use cases for Idris and if that presents a limitation other than being less popular than Rust.}

% Similar to Jif,
Flow Caml extends Objective Caml with a type system to track variable secrecy levels without requiring programmers to annotate the source code~\cite{FlowCAML}.
% While Flow Caml has full type inference, it does not support many of the object-oriented features of Objective Caml, such as polymorphic variables or labels.
Chapman and Hilton similarly extended SPARK Ada and the SPARK Examiner to include variable annotations~\citeyearpar{spark}. This extension does not support concurrency, and their declassification mechanism requires hiding the body of the declassifying subprogram from the Examiner.

To add static type-based IFC to an imperative language, prior work extends the language and requires a modified or custom compiler. Jif
% (along with its predecessor JFlow)
is an extension of Java that provides static type-based IFC~\cite{JFlow,Jif,ML_97}. It introduces types that represent secrecy and integrity labels, which programs use to annotate variables and expressions. Jif provides a compiler for the Jif language that checks types and flows, including implicit flows, and produces pure Java code with the same functionality as the original program.
Jif includes a variety of features such as support for dynamic labels and automatic inference. Follow-on research has extended Jif to solve a wide variety of IFC-related problems, such as erasure policies and reactive information flow~\cite{chong_dissertation, JRIF}.
%(cf.~\cite{???} \mike{Can we cite a few more papers from the Jif project?}\ada{\cite{chong_dissertation} extends Jif to incorporate decentralized label model with declassification, erasure policies, noninterference enforcement, decentralized robustness), \cite{JRIF} extends Jif to incorporate a reactive information flow automaton which specifies how transforming a value alters how the result might be used}. These features are outside the scope of this paper, which focuses on the potential for static typing--based IFC using an unmodified language and compiler.
The key drawback of Jif is that it requires programs to be written in a nonstandard language and to be processed with a nonstandard compiler, hampering adoption.
% In general, these properties will thwart adoption for most applications.

Volpano, Smith, et al.\ developed a model for IFC type systems in imperative languages that they later extended to support multithreading and non-interference, but (as far as we know) this model has not been implemented~\citeyearpar{IFC_multithreaded, SV_type_model}.

\iffalse
See \autoref{tbl:comparison} for a qualitative comparison of each of these solutions against \ifclib. 
\begin{table*}
    \begin{tabular}{|r|p{1.5cm}|p{1.5cm}|p{1.5cm}|p{1.5cm}|p{1.5cm}|p{1.5cm}|}
        \hline
        & \textbf{Target Language} & \textbf{Static IFC Support} & \textbf{Dynamic IFC Support} & \textbf{Uses Target Language} & \textbf{Uses Standard Compiler} & \textbf{Supports All Language Features} \\
        \hline
        \textbf{JFlow} & Java & \checkmark & \checkmark & \tickNo & \tickNo & \tickNo \\
        \hline
        \textbf{Java PDGs} & Java & \checkmark & \checkmark & \checkmark & \tickNo & \checkmark \\
        \hline
        \textbf{DepSec} & Idris & \checkmark & \tickNo & \checkmark & \checkmark & \checkmark \\
        \hline
        \textbf{Austin/Flanagan} & Java & \tickNo & \checkmark & \checkmark & \tickNo & \tickNo\\
        \hline
        \textbf{Laminar} & Java & \tickNo & \checkmark & \checkmark & \tickNo & \checkmark \\
        \hline
        \textbf{Co-Inflow} & Java & \tickNo & \checkmark & \tickNo & \tickNo & \tickNo \\
        \hline
        \textbf{\Ifclib} & Rust & \checkmark & \tickNo & \checkmark & \checkmark & \checkmark \\
        \hline
    \end{tabular}
    \caption{Comparison of other IFC tools against \ifclib.} \ada{Does it make sense to include SAILS here? SAILS doesn't have a target language (generic) which makes it difficult to answer whether it uses the target language, standard compiler, or supports all features. The answer I guess is yes because you can use almost any unmodified mainstream language.}
    \label{tbl:comparison}
\end{table*}
\fi

\label{subsec:rlbox}
RLBox is an approach for sandboxing libraries to prevent cross-library leaks~\cite{rlbox}. It uses a combination of static type-based analysis and run-time isolation to prevent data and control from transferring between libraries. RLBox, which targets C++ libraries,
% is a type-based approach for C++ that sandboxes libraries to prevent data- and control-flow leaks across libraries~\cite{rlbox}. RLBox uses software-based fault isolation and multi-process core isolation to prevent data or control from transferring between the sandbox and external libraries.
cannot provide a guarantee of side effect freedom, but rather requires programmers to write a validation check when handling tainted values. RLBox adds run-time and space overheads, unlike \ifclib. Another difference is that \ifclib's secrecy labels allow a hierarchy of labels, while RLBox has only binary labels: A value is ``tainted'' or it is not.

%\mike{TODO: Cite \cite{curricle}, which uses procedural macros to enforce a type system for idempotence.}
% similarities between Cocoon & Curricle: on top of Rust compiler, procedural macros, flow analysis, lattices, programmer annotations
% differences (Cocoon vs. Curricle): non-interference vs. idempotence (non-interference is more general), secrecy vs. integrity-style taint analysis, new type vs. attribute annotations, more or less ad hoc vs. formal model + core calculus + proof
In concurrent work, Curricle provides a type system for ensuring noninterference, in order to ensure region idempotence for intermittent programs~\cite{curricle}. Like \ifclib, Curricle employs Rust procedural macros to ensure that only well-typed programs can compile. Unlike \ifclib, Curricle uses procedural macros to perform type analysis at the syntactic level; in contrast, \ifclib transforms the region's code so that the compiler performs the type analysis. As a result, Curricle requires idempotent regions to be functions, and it cannot support unbounded loops or recursion in idempotent functions. Furthermore, Curricle is unsound with respect to syntactically invisible side effects such as implicit destructor calls, dereference operations, and overloaded operator calls (\autoref{sec:application-defined-functions-invisible}) since it operates at the syntactic level. The paper proves noninterference with respect to a formal model, not the Rust language or the Curricle implementation~\cite{curricle}.
% Where \ifclib focuses on IFC for secrecy, Curricle uses IFC for integrity control.
%\mike{Revised paragraph. Got a response from first Curricle first author agreeing that Curricle doesn't handle syntactically invisible side effects.}

\iffalse
\subsection{The Rust Programming Language}

%\mike{I think this Rust blurb is out of place and in any case can be removed considering the audience. When first mentioning Rust in Intro, text could mention its increasing popularity with a few words and a citation. I think \autoref{sec:design-impl} already talks about some of Rust's unique features; it's not clear we should talk about those before then.}
Rust was designed by Graydon Hoare at Mozilla Research in 2010 \cite{Hoare_Rust_Blog, Mozilla_Rust} and had its first stable release in 2015 \cite{Rust_GitHub}. Since then it has had 95 releases, over 3,500 contributors, and has held the title of Stack Overflow's ``most loved language'' by developers each year since 2016 \cite{stackOverflow_survey}. Unlike many other programming languages, Rust provides the low-level programming idioms often seen in C/C++ as well as strong safety guarantees found in high-level languages such as Java \cite{JJKD_2021}. Specifically, Rust provides type safety, memory safety, thread safety, a unique ownership model, and procedural macros (\`{a} la LISP).
% These features enable a static IFC solution without having to modify the Rust language or compiler.
% \mike{This sentence (inaccurately) makes the problem seem too easy.}
\fi
    
    % recap and additional discussion of "so what?" if necessary
    \section{Conclusion}
%\ada{TODO: Update to accommodate for the added limitations and future work section}

\Ifclib is the first static type-based IFC approach for an imperative language that uses the standard language and compiler.
% It represents an important advance in the state of the art for providing practical, language-level IFC for mainstream, high-performance languages.
% \Ifclib is a static type-based IFC library for the standard (unmodified) Rust language.
% that works with the standard Rust compiler.
The key technical contribution of \ifclib is its establishment of an effect system that tracks whether Rust functions have side effects. As a result, \ifclib allows programmers to define and perform arbitrary operations on secret data while ensuring that these operations do not violate IFC rules, all in pure Rust. \Ifclib can be incrementally deployed on existing systems without significantly affecting performance.

\ignore{Our evaluation on real applications and performance benchmarks shows that using \ifclib adds
% We retrofitted two real Rust applications, Spotify TUI and Servo, to use \ifclib to protect a secret value in each program. We found that the changes introduced
low compile-time overhead and no significant run-time overhead.}
% The changes were generally limited to the parts of the application dealing with the secret value.
% To better understand \ifclib's performance impact, we evaluated nine performance-oriented benchmarks transformed in their entirety by \ifclib, and found that \ifclib has no discernible impact on run time or memory usage and modest impacts on executable size, but does increase compile time noticeably.

\iffalse
In case studies on two real Rust applications, we showed
that relatively few modifications had to be made to the source code and no significant compile- or run-time overhead was incurred (less than 2\% in each case). \ada{I'm not super happy with using 2\% here - the percentage seems out of context and potentially leading to more questions since it's so general. Thoughts?}
%\ada{Add summary of Spotify TUI results}.
%In the Spotify TUI case study, we were able to integrate \ifclib with an increase of 1.67\% in lines of code, 1.78\% compile-time overhead, and no significant increase in compile-time. \ada{Better percentage wise? Also - checking I did this right: compile time was calculated as (Cocoon-original)/original*100. Meaning I didn't include the +/- variance. LOC was similar.} %of approximately 3 seconds (out of 145.9) \mike{Overhead matters as a percentage, i.e., the parenthetical is as important as the rest.} and did not significantly impact the run time. \ada{Add sentence about testing (e.g. number of tests) to match Servo summary.}
%To retrofit \ifclib into Servo, Mozilla's browser engine, we modified \ada{(todo - add \% modifications)} of the source code and incurred compile and run time overheads of 0.02\% and 0.01\%, respectively. %only 229 lines of code needed to be modified, of which 150 lines of code belonged to the test-suite \ada{Potentially update to match Table 1}. %To compare the compile and run-time burden of \ifclib, we compiled Servo both with and without \ifclib 10 times and ran its test-suite 20 times and found no significant increase in its compile or run-times. 
We also found, through analysis of simple examples, that programs using \ifclib do not result in a larger compiled code size. Since the \lstinline{Secret} type has no size overhead, the Rust compiler is able to perform standard compiler optimizations so that the \ifclib program has the exact same x86-64 code generated as the non-\ifclib program.
\mike{Can this paragraph summarize the takeaway/meaning rather than restate the results and methodology?}\ada{Better?}\mike{Yeah, but rewrote.}
\fi

\iffalse
\Ifclib does not currently support dynamic labels which disallows any application that does not know a secrecy label's value until run-time. This limitation is a known failing of static analysis and would require a run-time analysis solution and overhead to resolve. Extending \ifclib to support dynamic labels would broaden the possible use cases for the library.

Another area of possible improvement for \ifclib is to create auto-derivation for the lattice. Currently, \ifclib supports any lattice structure, but this structure is hardcoded into the library itself. Allowing the user to provide the base secrecy levels (i.e., those which are one ``step'' above public and have a secrecy label set with one element) would make the library more flexible. The challenge with this improvement is that the secrecy label inputs to the library have to be done in such a way that the derivation of the lattice is done before run time. 
\fi

    % data availability statement
    \section*{Data-Availability Statement}

    An artifact reproducing this paper's results is publicly available~\cite{cocoon-artifact}.
    %The \ifclib source code is also available on GitHub.\footnote{\url{https://github.com/PLaSSticity/Cocoon-implementation}}

% Note that indenting the acks causes a build error (acmart inexplicably uses specialcomment for the acks environment)
\begin{acks}
We thank the anonymous reviewers for valuable feedback,
Ethar Qawasmeh for help in the early stages of the project,
Chris Xiong for help with code and ideas, and
Chujun Geng for helpful feedback and discussions.
This work is supported by NSF grants CSR-2106117, XPS-1629126, and CNS-2207202.
\end{acks}

% \appendix
% \input{appendix}
% Note: supplementary material is not part of the paper and will need to be submitted as a separate document. Also reviewers may choose to ignore it.

    % references
    \bibliographystyle{ACM-Reference-Format}
    \bibliography{paper}

\end{document}